\documentstyle[12pt]{article}
\newcommand{\nc}{\newcommand}
\nc{\renc}{\renewcommand}

\nc{\etal}{\mbox{\it et al.}}
\nc{\ie}{{\it i.e.}}
\nc{\eg}{{\it e.g.}}

\renc{\thefootnote}{\arabic{footnote}}
\nc{\capt}[1]{{\bf Figure.} {\small\sl #1}}

\nc{\eqs}[2]{\mbox{Eqs.~(\ref{#1},\,\ref{#2})}}
\nc{\eq}[1]{\mbox{Eq.~(\ref{#1})}}

\nc{\figs}[2]{\mbox{Figs.~(\ref{#1},\,\ref{#2})}}
\nc{\fig}[1]{\mbox{Fig~.(\ref{#1})}}

\nc{\tag}[1]{\label{#1} \marginpar{{\footnotesize #1}}}
\nc{\mtag}[1]{\label{#1} \mbox{\marginpar{{\footnotesize #1}}}}
\renc{\baselinestretch}{1.2}
\newlength{\overeqskip}
\newlength{\undereqskip}
\nc{\be}[1]{\begin{equation} \mbox{$\label{#1}$}}
\nc{\bea}[1]{\begin{eqnarray} \mbox{$\label{#1}$}}
\nc{\Section}[2]{\section{#2}\label{#1}}
\nc{\Bibitem}[1]{\bibitem{#1}}
\nc{\Label}[1]{\label{#1}}

\nc{\eea}{\vspace{\undereqskip}\end{eqnarray}}
\nc{\ee}{\vspace{\undereqskip}\end{equation}}
\nc{\bdm}{\begin{displaymath}}
\nc{\edm}{\end{displaymath}}
\nc{\dpsty}{\displaystyle}
\nc{\bc}{\begin{center}}
\nc{\ec}{\end{center}}
\nc{\ba}{\begin{array}}
\nc{\ea}{\end{array}}
\nc{\bab}{\begin{abstract}}
\nc{\eab}{\end{abstract}}
\nc{\btab}{\begin{tabular}}
\nc{\etab}{\end{tabular}}
\nc{\bit}{\begin{itemize}}
\nc{\eit}{\end{itemize}}
\nc{\ben}{\begin{enumerate}}
\nc{\een}{\end{enumerate}}
\nc{\bfig}{\begin{figure}}
\nc{\efig}{\end{figure}}
\nc{\arreq}{&\!=\!&}
\nc{\arrmi}{&\!-\!&}
\nc{\arrpl}{&\!+\!&}
\nc{\arrap}{&\!\!\!\approx\!\!\!&}
\nc{\non}{\nonumber}
\nc{\align}{\!\!\!\!\!\!\!\!&&}

\def\lsim{\; \raise0.3ex\hbox{$<$\kern-0.75em
      \raise-1.1ex\hbox{$\sim$}}\; }
\def\gsim{\; \raise0.3ex\hbox{$>$\kern-0.75em
      \raise-1.1ex\hbox{$\sim$}}\; }
\nc{\DOT}{\hspace{-0.08in}{\bf .}\hspace{0.1in}}
\nc{\Laada}{\hbox {$\sqcap$ \kern -1em $\sqcup$}}
\nc\loota{{\scriptstyle\sqcap\kern-0.55em\hbox{$\scriptstyle\sqcup$}}}
\nc\Loota{{\sqcap\kern-0.65em\hbox{$\sqcup$}}}
\nc\laada{\Loota}
\nc{\qed}{\hskip 3em \hbox{\BOX} \vskip 2ex}

\nc{\real}{{\rm I \! R}}
\nc{\Z}{{\sf Z \!\!\! Z}}
\nc{\complex}{{\rm C\!\!\! {\sf I}\,\,}}
\def\bigid{\leavevmode\hbox{\small1\kern-3.8pt\normalsize1}}
\def\id{\leavevmode\hbox{\small1\kern-3.3pt\normalsize1}}
\nc{\slask}{\!\!\!/}
\nc{\bis}{{\prime\prime}}
\nc{\pa}{\partial}
\nc{\na}{\nabla}
\nc{\ra}{\rangle}
\nc{\la}{\langle}
\nc{\goto}{\rightarrow}
\nc{\swap}{\leftrightarrow}

\nc{\EE}[1]{ \mbox{$\cdot10^{#1}$} }
\nc{\abs}[1]{\left|#1\right|}
\nc{\at}[2]{\left.#1\right|_{#2}}
\nc{\norm}[1]{\|#1\|}
\nc{\abscut}[2]{\Abs{#1}_{\scriptscriptstyle#2}}
\nc{\vek}[1]{{\rm\bf #1}}
\nc{\integral}[2]{\int\limits_{#1}^{#2}}
\nc{\inv}[1]{\frac{1}{#1}}
\nc{\dd}[2]{{{\partial #1}\over{\partial #2}}}
\nc{\ddd}[2]{{{{\partial}^2 #1}\over{\partial {#2}^2}}}
\nc{\dddd}[3]{{{{\partial}^2 #1}\over
	{\partial #2 \partial #3}}}
\nc{\dder}[2]{{{d #1}\over{d #2}}}
\nc{\ddder}[2]{{{d^2 #1}\over{d {#2}^2}}}
\nc{\dddder}[3]{{d^2 #1}\over
	{d #2 d #3}}
\nc{\dx}[1]{d\,^{#1}x}
\nc{\dy}[1]{d\,^{#1}y}
\nc{\dz}[1]{d\,^{#1}z}
\nc{\dl}[1]{\frac{d\,^{#1}l}{(2\pi)^{#1}}}
\nc{\dk}[1]{\frac{d\,^{#1}k}{(2\pi)^{#1}}}
\nc{\dq}[1]{\frac{d\,^{#1}q}{(2\pi)^{#1}}}

\nc{\cc}{\mbox{$c.c.$ }}
\nc{\hc}{\mbox{$h.c.$ }}
\nc{\cf}{cf.\ }
\nc{\erfc}{{\rm erfc}}
\nc{\Tr}{{\rm Tr\,}}
\nc{\tr}{{\rm tr\,}}
\nc{\pol}{{\rm pol}}
\nc{\sign}{{\rm sign}}
\nc{\bfT}{{\bf T }}

\nc{\cA}{{\cal A}}
\nc{\cB}{{\cal B}}
\nc{\cD}{{\cal D}}
\nc{\cE}{{\cal E}}
\nc{\cG}{{\cal G}}
\nc{\cH}{{\cal H}}
\nc{\cL}{{\cal L}}
\nc{\cO}{{\cal O}}
\nc{\cT}{{\cal T}}
\nc{\cN}{{\cal N}}
\nc{\rvac}[1]{|{\cal O}#1\rangle}
\nc{\lvac}[1]{\langle{\cal O}#1|}
\nc{\rvacb}[1]{|{\cal O}_\beta #1\rangle}
\nc{\lvacb}[1]{\langle{\cal O}_\beta #1 |}
\nc{\bb}{\bar{\beta}}
\nc{\bt}{\tilde{\beta}}
\nc{\ctH}{\tilde{\cal H}}
\nc{\chH}{\hat{\cal H}}
\nc{\1}{\aa}
\nc{\2}{\"{a}}
\nc{\3}{\"{o}}
\nc{\4}{\AA}
\nc{\5}{\"{A}}
\nc{\6}{\"{O}}
\nc{\al}{\alpha}
\nc{\g}{\gamma}
\nc{\Del}{\Delta}
\nc{\e}{\epsilon}
\nc{\eps}{\epsilon}
\nc{\lam}{\lambda}
\nc{\Om}{\Omega}
\nc{\ve}{\varepsilon}
\nc{\mn}{{\mu\nu}}
\nc{\k}{\kappa}
\nc{\vp}{\varphi}

\nc{\advp}[3]{{\it  Adv.\ in\ Phys.\ }{{\bf #1} {(#2)} {#3}}}
\nc{\annp}[3]{{\it  Ann.\ Phys.\ (N.Y.)\ }{{\bf #1} {(#2)} {#3}}}
\nc{\apl}[3]{{\it  Appl. Phys. Lett. }{{\bf #1} {(#2)} {#3}}}
\nc{\apj}[3]{{\it  Ap.\ J.\ }{{\bf #1} {(#2)} {#3}}}
\nc{\apjl}[3]{{\it  Ap.\ J.\ Lett.\ }{{\bf #1} {(#2)} {#3}}}
\nc{\app}[3]{{\it Astropart.\ Phys.\ }{{\bf #1} {(#2)} {#3}}}  
\nc{\cmp}[3]{{\it  Comm.\ Math.\ Phys.\ }{{ \bf #1} {(#2)} {#3}}}
\nc{\cqg}[3]{{\it  Class.\ Quant.\ Grav.\ }{{\bf #1} {(#2)} {#3}}}
\nc{\epl}[3]{{\it  Europhys.\ Lett.\ }{{\bf #1} {(#2)} {#3}}}
\nc{\ijmp}[3]{{\it Int.\ J.\ Mod.\ Phys.\ }{{\bf #1} {(#2)} {#3}}}
\nc{\ijtp}[3]{{\it Int.\ J.\ Theor.\ Phys.\ }{{\bf #1} {(#2)} {#3}}}
\nc{\jmp}[3]{{\it  J.\ Math.\ Phys.\ }{{ \bf #1} {(#2)} {#3}}}
\nc{\jpa}[3]{{\it  J.\ Phys.\ A\ }{{\bf #1} {(#2)} {#3}}}
\nc{\jpc}[3]{{\it  J.\ Phys.\ C\ }{{\bf #1} {(#2)} {#3}}}
\nc{\jap}[3]{{\it J.\ Appl.\ Phys.\ }{{\bf #1} {(#2)} {#3}}}
\nc{\jpsj}[3]{{\it J.\ Phys.\ Soc.\ Japan\ }{{\bf #1} {(#2)} {#3}}}
\nc{\lmp}[3]{{\it Lett.\ Math.\ Phys.\ }{{\bf #1} {(#2)} {#3}}}
\nc{\mpl}[3]{{\it  Mod.\ Phys.\ Lett.\ }{{\bf #1} {(#2)} {#3}}}
\nc{\ncim}[3]{{\it  Nuov.\ Cim.\ }{{\bf #1} {(#2)} {#3}}}
\nc{\np}[3]{{\it  Nucl.\ Phys.\ }{{\bf #1} {(#2)} {#3}}}
\nc{\pr}[3]{{\it Phys.\ Rev.\ }{{\bf #1} {(#2)} {#3}}}
\nc{\pra}[3]{{\it  Phys.\ Rev.\ A\ }{{\bf #1} {(#2)} {#3}}}
\nc{\prb}[3]{{\it  Phys.\ Rev.\ B\ }{{{\bf #1} {(#2)} {#3}}}}
\nc{\prc}[3]{{\it  Phys.\ Rev.\ C\ }{{\bf #1} {(#2)} {#3}}}
\nc{\prd}[3]{{\it  Phys.\ Rev.\ D\ }{{\bf #1} {(#2)} {#3}}}
\nc{\prl}[3]{{\it Phys\ Rev.\ Lett.\ }{{\bf #1} {(#2)} {#3}}}
\nc{\pl}[3]{{\it  Phys.\ Lett.\ }{{\bf #1} {(#2)} {#3}}}
\nc{\prep}[3]{{\it Phys\. Rep.\ }{{\bf #1} {(#2)} {#3}}}
\nc{\prsl}[3]{{\it Proc.\ R.\ Soc.\ London\ }{{\bf #1} {(#2)} {#3}}}
\nc{\ptp}[3]{{\it  Prog.\ Theor.\ Phys.\ }{{\bf #1} {(#2)} {#3}}}
\nc{\ptps}[3]{{\it  Prog\ Theor.\ Phys.\ suppl.\ }{{\bf #1} {(#2)} {#3}}}
\nc{\physa}[3]{{\it  Physica\ A\ }{{\bf #1} {(#2)} {#3}}}
\nc{\physb}[3]{{\it  Physica\ B\ }{{\bf #1} {(#2)} {#3}}}
\nc{\phys}[3]{{\it Physica\ }{{\bf #1} {(#2)} {#3}}}
\nc{\rmp}[3]{{\it  Rev.\ Mod.\ Phys.\ }{{\bf #1} {(#2)} {#3}}}
\nc{\rpp}[3]{{\it Rep.\ Prog.\ Phys.\ }{{\bf #1} {(#2)} {#3}}}
\nc{\sjnp}[3]{{\it Sov.\ J.\ Nucl.\ Phys.\ }{{\bf #1} {(#2)} {#3}}}
\nc{\spjetp}[3]{{\it Sov.\ Phys.\ JETP\ }{{\bf #1} {(#2)} {#3}}}
\nc{\yf}[3]{{\it Yad.\ Fiz.\ }{{\bf #1} {(#2)} {#3}}}
\nc{\zetp}[3]{{\it Zh.\ Eksp.\ Teor.\ Fiz.\  }{{\bf #1}  {(#2)} {#3}}}
\nc{\zp}[3]{{\it Z.\ Phys.\ }{{\bf #1} {(#2)} {#3}}}
\nc{\ibid}[3]{{\sl ibid.\ }{{\bf #1} {#2} {#3}}}
\nc{\rf}[1]{(\ref{#1})}
\nc{\nn}{\nonumber \\*}
\nc{\bfB}{\bf{B}}
\nc{\bfv}{\bf{v}}
\nc{\bfx}{\bf{x}}
\nc{\bfy}{\bf{y}}
\nc{\vx}{\vec{x}}
\nc{\vy}{\vec{y}}
\nc{\oB}{\overline{B}}
\nc{\oI}{\overline{I}}
\nc{\oR}{\overline{R}}
\nc{\rar}{\rightarrow}
\nc{\ti}{\times}
\nc{\slsh}{\hskip-5pt/}
\nc{\sm}{Standard~Model~}
\nc{\MP}{M_{\rm Pl}}
\nc{\tp}{t_{\rm Pl}}
\nc{\ave}{\bar{E}}

\renc{\min}{p_{\rm min}}
\renc{\max}{p_{\rm max}}
\nc{\pmin}{p_{\rm min}}
\nc{\pmax}{p_{\rm max}}
\nc{\fo}{f_0}
\nc{\foi}{f_{0,i}\,}
\nc{\fop}{f_0^P}
\nc{\fou}{f_0^U}
\def\sepand{\rule{14cm}{0pt}\and}
\nc{\eff}{{\rm eff}}
\nc{\MT}{M_{\rm T}}
\nc{\ML}{M_{\rm L}}
\nc{\kk}{\vek{k}}
\nc{\pp}{{\rm p}}
\nc{\pt}{\partial_t}
\nc{\half}{{1\over 2}}
\nc{\w}{\omega}
\nc{\uhat}{\hat{U}_\w}

\begin{document}

{\title{{\hfill {{\small  TURKU-FL-P33-99 
        }}\vskip 1truecm}
{\bf Analytical and numerical properties of Q-balls}}

 
\author{
{\sc Tuomas Multam\" aki$^{1}$}\\
{\sl and}\\
{\sc Iiro Vilja$^{2}$ }\\ 
{\sl Department of Physics,
University of Turku} \\
{\sl FIN-20014 Turku, Finland} \\
\sepand
}
\maketitle}
\vspace{2cm}
\begin{abstract}
\noindent 
Stable non-topological solitons, Q-balls, are studied using analytical
and numerical methods. Three different physically interesting 
potentials that support Q-ball solutions are considered: two typical
polynomial potentials and a logarithmic potential inspired by supersymmetry. 
It is shown that Q-balls in
these potentials exhibit different properties in the thick-wall limit
where the charge of a Q-ball is typically considerably smaller
than in the thin-wall limit.
Analytical criteria are derived to check whether stable Q-balls exists
in the thick-wall limit for typical potentials. 
Q-ball charge, energy and profiles are presented for each
potential studied.
Evaporation rates are calculated in the perfect thin-wall limit 
and for realistic Q-ball profiles. It is shown that in each case 
the evaporation rate increases with decreasing charge.
\end{abstract}
\vfill
\footnoterule
{\small$^1$tuomas@maxwell.tfy.utu.fi,  $^2$vilja@newton.tfy.utu.fi}
\thispagestyle{empty}
\newpage
\setcounter{page}{1}
\section{Introduction}
A scalar field theory with a spontaneously broken $U(1)$-symmetry
may contain stable non-topological solitons \cite{leepang}, Q-balls
\cite{coleman}. 
A Q-ball is a coherent state of a complex scalar field
that carries a global $U(1)$ charge. 
(One should be careful not to use the word soliton by itself
to describe these objects since Q-balls are not solitons in the
strict sense of the word).
In the sector of fixed charge
the Q-ball solution is the ground state so that its stability and 
existence are due to the conservation of the $U(1)$ charge.
Q-balls associated with a local $U(1)$ charge have also been
constructed \cite{klee} but these will not be discussed in this paper.

In realistic theories Q-balls are generally allowed in supersymmetric 
generalizations of the standard model with flat directions in their
scalar potentials. Q-balls can be formed from scalars carrying
a conserved $U(1)$ charge.
Flat directions in the Minimally Supersymmetric Standard Model (MSSM)
\cite{nilles}
have been shown to exist and they have
been classified in \cite{dine}.
In particular Q-balls have been shown to be present in 
the MSSM
\cite{kusenko2,enqvist1} 
where leptonic and baryonic balls may exist.
For a Q-ball solution to be possible in the scalar sector of the theory
with potential $U(\phi)$,
$U(\phi)/\phi^2$ should have a global minimum at a non zero value of
$\phi$ \cite{coleman}. Such renormalizable $F$- and $D$-flat directions 
where this is possible
in the MSSM are for example the $H_u^0L$-direction (where
the expectation values $H_u^0$ and $\nu_L$ are non-zero and equal)
and the $u^cd^cd^c$-direction (along which $u_c$, $d_c$ and $d'_c$
VEVs are non-zero) \cite{enqvist1}. For a large $\phi$ a D-flat direction
in the potential (suppressed at the Planck scale) 
is of the form \cite{enqvist1}
\be{dflat}
U(\phi)=m_S^2|\phi|^2+{\lambda^2|\phi|^{2(d-1)}\over M_{Pl}^{2(d-3)}}+
({A_\lambda\lambda\phi^d\over M_{Pl}^{d-3}}+\hc),
\ee
where $m_S$ is the soft SUSY breaking mass parameter, $A_\lambda$
the $A$-term and $d$ is the dimension of the non-renormalizable
term in the superpotential. For Q-balls to exist in the theory,
$m_S$ must be $\phi$ dependent and furthermore be such that
$U(\phi)/|\phi|^2$ has a global minimum at a non-zero $\phi$.

Q-balls may be formed in the early universe by a mechanism
\cite{enqvist1,kusenko3}
that is
closely related to the Affleck-Dine baryogenesis \cite{adb}.
A scalar condensate first forms along a flat direction in the potential.
In the MSSM this is a D-flat direction composed of squark and possibly 
slepton fields. The spatially homogeneous condensate then develops an
instability that leads to the formation of areas of different charge
densities. This charge pattern can then lead to the creation of Q-balls
(recent work \cite{qaxitons} shows that initially the objects formed
from the AD-condensate are not in general Q-balls, but Q-axitons).
In the MSSM Q-balls formed in such a way are baryon (and lepton)
number carrying balls with charges typically larger than 
$10^{14}$ \cite{enqvist1}. 
The subsequent evolution of the Q-balls then critically depends upon the 
mechanism how supersymmetry is broken in the theory \cite{enqvistdm}. 
If SUSY breaking occurs via a gauge mediated mechanism \cite{dimo}
the potential for the scalar condensate is completely flat
($m_S$ behaves as $\phi^{-2}$ for large enough $\phi$ in this scenario)
and the resulting Q-balls can have huge charges \cite{kusenko3,
qpower}. For a large enough
Q-ball with no lepton number, all decay modes can be excluded and the Q-ball can be completely 
stable. 
If, however, the SUSY breaking mechanism is via a supergravity
hidden sector \cite{nilles} the scalar potential is not completely flat
but radiative corrections allow for Q-ball solutions to exist
in the potential.
The mass parameter
$m_S$ is in this case given by \cite{enqvist1}
\be{msbeh}
m_S^2\approx m_o^2(1+K \log({|\phi|^2\over M_X^2})),
\ee
where $K$ is typically $-0.01$ - $-0.1$ and $M_X$ is a large mass scale
($m_S^2(M_X)=m_o$).
In this case the Q-balls are generally unstable and can typically
decay to quarks and nucleons \cite{enqvistdm}.

The possible existence of Q-balls has inspired new ideas and 
particularly their cosmological significance has been
discussed recently. The question of Q-ball stability plays an
important role in their importance to cosmology.
If stable Q-balls are formed in the early universe they can 
contribute to the dark matter content of the universe \cite{kusenko3,
kusennko2}. These can be huge balls with charges of the order of
$10^{20}$ \cite{kusenko3} but also very small Q-balls can be considered
as dark matter \cite{kusennko2}.
Stable Q-balls can also catalyze explosions of neutron stars
by accumulating inside the stars and decreasing the mass
of the star by absorbing baryons \cite{wreck}.
Decaying Q-balls can also be of crucial cosmological significance.
If Q-balls decay after the electroweak phase transition, they can 
protect baryons from the erasure of baryon number due
to sphaleron transitions \cite{enqvistdm}. Possibly they can also
create the asymmetry from a $B-L$ conserving condensate \cite{enqvist1}.
Furthermore the Q-ball decay may result in the production of dark matter
in the form of the lightest supersymmetric particle (LSPs). This process 
can then naturally explain the baryon to dark matter ratio of the universe 
\cite{enqvistdm}.
Small Q-balls produced in a collider would also 
have interesting experimental properties \cite{kusennko2}.
It has even been suggested that Q-ball catalyzed processes may provide
us with an inexhaustible source of energy \cite{qpower}.

The detection of Q-balls was discussed in \cite{qdet} where
detection was considered for two types of Q-balls, 
Supersymmetric Electrically Neutral Solitons (SENS) and
Supersymmetric Electrically Charged Solitons (SECS). The mode
of detection depends crucially on the possible electric charge 
carried by the Q-ball. Assuming that all or most of dark matter is
in Q-balls, the number density of stable Q-balls was estimated to be 
\cite{qdet}
\be{qdensity}
n_Q\sim 5\times 10^{-5} Q_B^{-3/4}({1\ \textrm{TeV}\over m})
\ \textrm{cm}^{-3},
\ee
where $Q_B$ is the charge of the baryonic ball and 
$m$ is the mass of the baryon number carrying scalar. 
Further assuming that
the average velocity of Q-balls is $\sim 10^{-3}c$, experimental limits
can then be calculated. From the MACRO search \cite{macro} a lower bound
on SECS charge is $\sim 10^{21}$. From monopole searches \cite{baikal}
the lower bound for SENS charge is $\sim 10^{22}$.
In considering these limits one must bear in mind that here it has
been assumed that dark matter consist mostly of Q-balls. If
there are far fewer Q-balls in the universe, however, then this
bounds can become significantly lower. Also if Q-balls are unstable
and evaporate there may be only indirect evidence of their 
existence present in the universe today.

In this paper some of the analytical and numerical properties of Q-balls 
are studied. We first 
derive some simple analytical results for Q-balls in generic scalar
potentials. Q-balls are then studied numerically in these typical
potentials. In the last part of this paper we 
consider Q-ball evaporation for realistic Q-ball profiles and 
present numerical results.

\section{Q-ball solutions}
We consider a field theory with a $U(1)$ symmetric scalar potential.
$U(\phi)$ has a global minimum $U(0)=0$ at $\phi=0$ and the
scalar field $\phi$ has a unit charge with respect to the $U(1)$.
The Lagrangian density for the theory is given by 
\be{qlagrangian}
\cL=\partial_\mu\phi^*\partial^\mu\phi-U(\phi).
\ee
The charge and energy of a given field configuration $\phi$ are 
\cite{kusenko2}
\be{charge}
Q={1\over 2i}\int (\phi^*\pt\phi-\phi\pt\phi^*)d^3x
\ee
\be{energy}
E=\int d^3x[\half\abs{\dot{\phi}}^2+\half\abs{\nabla\phi}^2+U(\phi)].
\ee

We wish to find the minimum energy configuration at a fixed charge,
\ie the Q-ball solution. If it is energetically favourable to store
charge in a Q-ball compared to free particles, the Q-ball will be stable
(in realistic theories, however, additional couplings may lead to 
evaporation in a different sector of the theory). 
Hence for a stable Q-ball, condition
$E<mQ$, where $m$ is the mass of the $\phi$-scalar, must hold.
Minimizing the energy
is done by using Lagrange multipliers \cite{kusenko2} \ie  we wish to minimize
\be{epsi}
\epsilon_\omega=E+\omega\big[ Q-{1\over 2i}\int (\phi^*\pt\phi-\phi\pt\phi^*)d^3x\big],
\ee
where $\omega$ is a Lagrange multiplier.
Separating (\ref{epsi}) into time and space dependent parts we get
\be{epsi2}
\epsilon_\omega=\int d^3x\half\abs{\pt\phi-i\w\phi}^2+\int d^3x\big[\half\abs{\nabla\phi}^2+U_\w(\phi)\big]+\w Q,
\ee
where
\be{uhat}
U_\w(\phi)=U(\phi)-\half \w^2\phi^2.
\ee
To minimize (\ref{epsi}) we first choose 
\be{sol1}
\phi(x,t)=e^{i\w t}\phi(x),
\ee
where $\phi(x)$ is now time independent and real.
Charge and energy are now given by
\be{Q2}
Q=\w\int \phi(x)^2 d^3x
\ee
and
\be{E2}
E=\int d^3x\big[\half\abs{\nabla\phi(x)}^2+U_\w(\phi(x))\big]+\w Q.
\ee
The problem of minimizing the energy for a fixed $\w$ in four
dimensions is then equivalent to the 
problem of finding the bounce for tunneling in $D=3$ \cite{kusenko2}. 
If a bounce solution exists it is spherically
symmetric \cite{cole2}. 
Hence the Q-ball solution also has a spherical symmetry,
\be{spher}
\phi(x)=\phi(r).
\ee
The charge and energy expressions for a spherically symmetric 
Q-ball are given by
\be{Q3}
Q=4\pi\w\int_0^\infty r^2 \phi(r)^2 dr
\ee
and
\be{E3}
E=4\pi\int_0^\infty \big[\half(\dd{\phi}{r})^2+U_\w(\phi(r))\big]r^2 dr+\w Q.
\ee
The equation of motion at a fixed $\w$ is 
\be{eom}
\ddder{\phi}{r}+{2\over r}\dder{\phi}{r}=\dd{U(\phi)}{\phi}-\w^2\phi.
\ee
To obtain the Q-ball profiles we must solve (\ref{eom}) with
boundary conditions $\phi'(0)=0,\ \phi(\infty)=0$.
The initial value of the field, $\phi(0)$, is left as a free parameter. 
This is, in general, a numerical problem that is discussed later in 
this paper.

Some properties of Q-ball solutions in general potentials
can be studied analytically. These are discussed in the following. 

\subsection{Behaviour of Q-ball charge as $\w\rightarrow m$}

We examine a general potential of the following form in $D$ dimensions,
\be{pot}
U(\phi)=\half (m^2-\w^2)\phi^2-\lambda \phi^A+\cO(\phi^{A+1}),
\ee
where $A>2$ and $\lambda$ is positive.
When $r$ is large the 'friction' term ${2\over r}\phi'$
is small so that the equation of motion can be approximated by
\be{eomap}
\phi''\approx \partial_\phi U_\w(\phi).
\ee
As $r\rightarrow\infty$, $\phi$ tends to zero and we can neglect
all but the effective mass term in the potential $U_\w(\phi)$ so that from
(\ref{eomap}) we get
\be{appvp}
\phi(r)\sim e^{-\sqrt{m^2-\w^2}r}.
\ee
Furthermore, since when $\w\rightarrow m$ the position of the zero of 
$U_\w(\phi)$ must be near the origin \cite{leepang}, 
we can improve the approximation (\ref{appvp}) to
\be{app2}
\phi\sim (m^2-\w^2)^{1\over A-2}e^{-\sqrt{m^2-\w^2} r}.
\ee
We can now calculate the charge at this limit,
\be{N}
Q\sim\lim_{\w\rightarrow m} \int d^D x\phi^2\sim(m^2-\w^2)^{2\over A-2}\int_0^\infty dr r^{D-1}
e^{-2\sqrt{m^2-\w^2}}\sim (m^2-\w^2)^{({2\over A-2}-{D\over 2})}.
\ee 
Clearly the value of $\eta\equiv ({2\over A-2}-{D\over 2})$ now determines
the behaviour of the Q-ball charge as $\w\rightarrow m$;
\be{nbehav}
\left\{
\begin{array}{lll}
\eta>0 & \Rightarrow & Q\goto 0\\
\eta=0 & \Rightarrow & Q\goto \textrm{const.}\\
\eta<0 & \Rightarrow & Q\goto \infty.
\end{array}
\right.
\ee
This result agrees with the results for $A=4$ presented in \cite{leepang}.

Even though solving the Q-ball equation of motion is generally a 
numerical problem, in certain cases analytical approximations can
be useful. Two limiting cases are discussed in the following,
the thin-wall and thick-wall approximations.

\subsection{Thin-wall approximation}
In a flat potential a Q-ball may have a large charge and the effect
of its surface can be ignored compared to the bulk of the ball.
Such a thin-walled Q-ball \cite{coleman} may be approximated by an ansatz
\be{thinwall}
\phi(r)=\left\{ \begin{array}{ll}
\phi_0, & r<R\\
0, & r\geq R,
\end{array}\right.
\ee
so that all surface effects are ignored.
$\phi_0$ is defined such that $U(\phi)/\phi^2$ is minimum for $\phi=\phi_0$.
The charge is given by
\be{thwcharge}
Q={4\pi\over 3}\w R^3 \phi_0^3=V\w\phi_0^3
\ee
and the energy is
\be{thwenergy}
E=Q\sqrt{2U(\phi_0)\over\phi_0^2}.
\ee
In a flat potential ratio $E/Q$ then decreases for increasing $\phi$
so that it becomes more and more energetically favourable to store 
quanta in the Q-ball.
This approximation is accurate for very large, thin-walled Q-balls. 
For the types of potentials studied in this paper,
a thin-wall approximation is valid for small $\w$ and it
generally deteriorates as $\w$ tends to $m$.
Generally the existence of a thin-walled Q-balls is associated with
completely flat directions in the scalar potential of the theory. 
These can lead to the formation of Q-balls with very large charges
\cite{kusenko3,enqvist1}.

\subsection{Thick-wall approximation}
As $\w$ tends to $m$, the Q-ball configuration becomes more and more
thick-walled so that at some point the thin-wall approximation 
is no longer valid.
When $\w$ is close to $m$ another approximation, the thick-wall
approximation, can be used.

\begin{figure}
\leavevmode
\centering
\vspace*{95mm}
\includegraphics{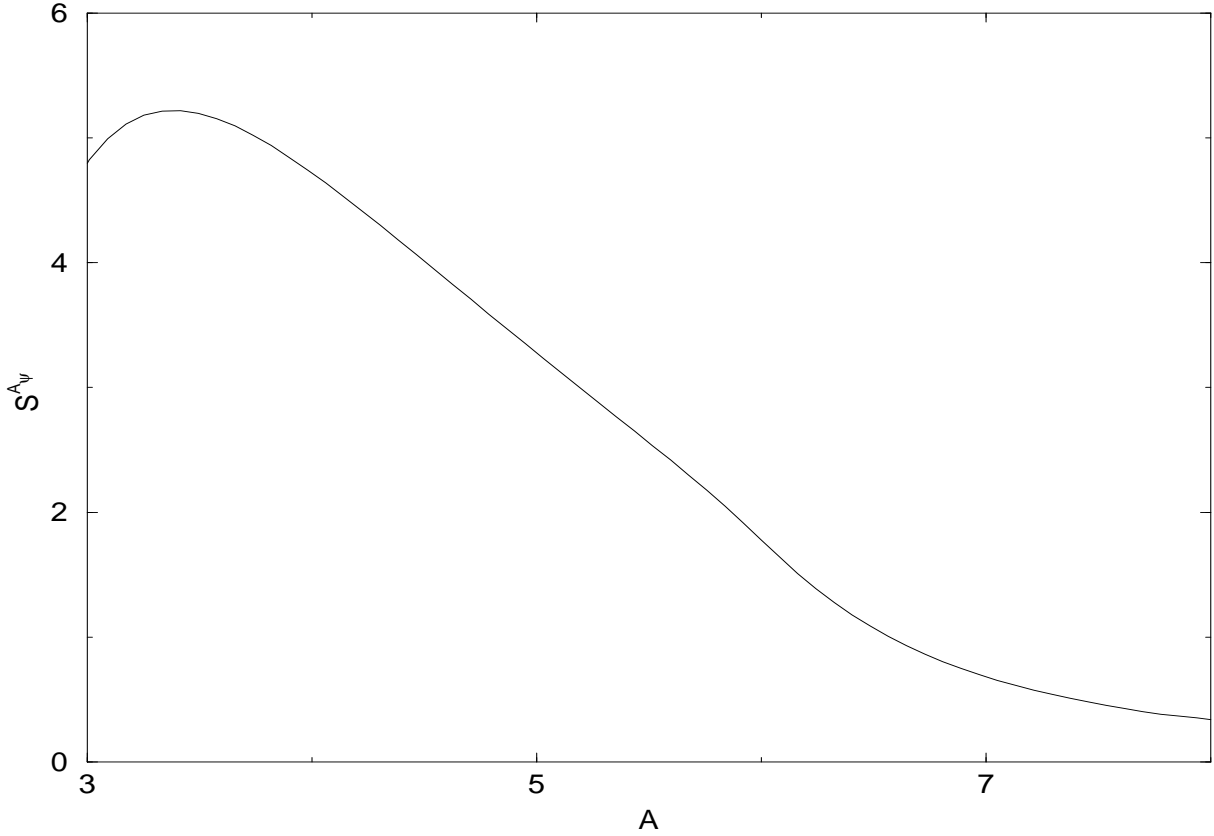}
\caption{The three dimensional bounce for different values of $A$}
\label{spsi}
\end{figure}

Kusenko \cite{kusenko2} considered a thick-wall approximation in the potential
\be{kpot}
U(\phi)=\half m^2\phi^2-\lambda\phi^3+\lambda'\phi^4,
\ee
so that 
\be{kpothat}
U_\w(\phi)=\half (m^2-\w^2)\phi^2-\lambda\phi^3+\lambda'\phi^4.
\ee
We generalize the potential to
\be{apot}
U(\phi)=\half m^2\phi^2-\lambda\phi^A+\cO(\phi^{A+1}),
\ee
where $A>2$ and $\lambda>0$
and consider the possibility of Q-balls in $D$ dimensions.
In the thick-wall limit, \ie  when the higher terms are negligible,
$\cO(\phi^{A+1})\sim 0$,
\be{apot2}
U_\w(\phi)\approx\half (m^2-\w^2)\phi^2-\lambda\phi^A.
\ee
Introducing dimensionless space-time coordinates, $\xi_i=(M^2-\w^2)^\half x_i$,
and a dimensionless field, $\psi=({\lambda\over M^2-\w^2})^{1\over A-2}\phi$,
the expression for energy (\ref{E2}) can be written as
\be{dimlesse}
\epsilon_\w=\lambda^{2\over 2-A}S_\psi^A (m^2-\w^2)^{2A-AD+2D\over 2(A-2)}+
\w Q,
\ee
where $S_\psi^A$ is the action of the bounce in the dimensionless
potential $\half\psi^2- \psi^A$. The exact value of the action is 
not essential to our argument 
provided that it is well behaved as $D$ and $A$ grow. $S_\psi^A$ is plotted
in Fig. \ref{spsi}. for $D=3$. We have also checked that $S_\psi^A$ is
well behaved for higher dimensions. 

Q-balls may exist provided that $\epsilon_\w$ has a minimum for some 
$0<\w_c<m$ such that $\epsilon_{\w_c}<mQ$. Furthermore for the
thick wall-approximation to be valid $\w_c$ must be close to $m$.

The expression for energy (\ref{dimlesse}) and its derivative 
\be{dedw}
\dd{\epsilon_\w}{\w}=Q-\lambda^{2\over 2-A}2\w(m^2-\w^2)^\zeta S_\psi^A (\zeta+1),
\ee
where $\zeta\equiv {4+2D-AD\over 2(A-2)}$,
exhibit different behaviour as $\w\rightarrow
m$, depending on the value of $\zeta$: 

If $\zeta>0$, $\epsilon_\w\rightarrow mQ$
and $\dd{\epsilon_\w}{\w}\rightarrow Q$. In this case 
Q-balls exist in the thick-wall limit provided that
the condition
\be{qcrit}
Q<2\lambda^{2\over 2-A}m^{2\zeta+1}(2\zeta)^\zeta
(1+2\zeta)^{-\zeta-{1\over 2}}S_{\psi}^A(\zeta+1)
\ee
holds.

If $\zeta=0$ no local minimum exists in the interval $0<\w<m$ and hence 
no stable Q-balls exist in the thick-wall limit.

If $-1<\zeta<0$, $\epsilon_\w\rightarrow mQ$ but now
$\dd{\epsilon_\w}{\w}\rightarrow -\infty$ so that there are no stable 
Q-ball solutions in the thick-wall limit.

If $\zeta=-1$ no local minimum exists and hence there are no stable
Q-ball solutions in the thick-wall limit.

If $\zeta<-1$, $\epsilon_\w\rightarrow\infty$ and 
$\dd{\epsilon_\w}{\w}\rightarrow \infty$. Again, there are no
stable Q-balls in the thick-wall limit.

To summarize, for stable Q-balls to exist in the thick-wall limit,
conditions
\bea{qconds}
\begin{array}{l}
0<4+2D-AD \\
Q<2\lambda^{2\over 2-A}m^{2\zeta+1}(2\zeta)^\zeta
(1+2\zeta)^{-\zeta-{1\over 2}}S_{\psi}^A(\zeta+1), 
\end{array}
\eea
must hold.

For example, when $D=3$, $A$ must be less than ${10\over 3}$. In
\cite{kusenko2} it was shown that Q-balls exist at the thick-wall limit when
$D=3,\ A=3$. When $D=3$, $A=4$ one can easily show that 
\be{d3a4}
E=mQ\sqrt{1+\beta^{-2}},
\ee
where $\beta={\lambda Q\over S_\psi^4}$. Since $\beta>0$, $E>mQ$ and 
hence there are no stable Q-ball solutions. This is also demonstrated
numerically in the following section.

\section{Numerical results}
We have examined three different types of potentials numerically, namely
\bea{potforms}
U_1(\phi) & = & {1\over 2} m_1^2\phi^2-\alpha_1\phi^4+\lambda_1\phi^6\\
U_2(\phi) & = & {1\over 2} m_2^2\phi^2-{\alpha_2\over 3}\phi^3+{\lambda_2\over 4}
\phi^4\\
U_3(\phi) & = & m_3^2\phi^2(1-K \log({\phi^2\over M^2}))+\lambda_3
\phi^{10}.
\eea
This choice of potentials is motivated by physical arguments.
$U_1(\phi)$ is a typical simple potential in field theories
that can contain extended objects of the Q-ball type. $U_2(\phi)$
is a typical potential in finite temperature field theories as the
the presence of the $\phi^3$ term indicates. $U_3(\phi)$ corresponds to
a D-flat direction in the MSSM with SUSY broken via
a conventional supergravity hidden sector.
The parameters are chosen for each potential such that 
when $\w=0$ there is a degenerate global minimum,
$U(\phi=0)=0$ and there exists a value $\phi_0>0$ such that
$U(\phi_0)$=0, which sets $\alpha_1=\sqrt{2m_1^2\lambda_1}$,
$\alpha_2=\sqrt{{9\over 2}m_2^2\lambda_2}$ and 
$M=({1\over 4}K m_3^2\lambda_3^{-1}\exp(-1-{4\over K}))^{1\over 8}$.
This choice ensures that there is no spontaneous breakdown
of $U(1)$-symmetry at $\omega=0$.
Furthermore the parameters must chosen appropriately to ensure
the existence of Q-balls as $\w$ increases \ie  the coefficient of
the highest power should be small enough compared to $m$.
We also need to ensure the correct behaviour at large $\phi$ \ie 
$U(\phi)\rightarrow\infty$ as $\phi\rightarrow\infty$.

The parameter values we have chosen to plot the figures are: 
\bea{numpots}
\begin{array}{lll}
m_1=1 & m_2=1 & m_3=100\ \textrm{GeV}\\
\lambda_1=0.001 & \lambda_2=0.01 & \lambda_3={1\over M_{Pl}^6}\\
K=0.1.
\end{array}
\eea
Note that only in potential $U_3(\phi)$ we need to introduce 
dimensional units due to the presence of the Planck mass
parameter. 
The choice of parameters is not special for $U_1(\phi)$ and 
$U_2(\phi)$, the value of $m$ simply 
sets the scale of the units, $\lambda_1$ and $\lambda_2$ set the 
profile of the
potential. The parameter values for $U_3(\phi)$ are typical
values for a realistic potential.
Solving for the place of the local minimum $\phi_1$ and 
maximum $\phi_2$, we get
\bea{scales}
\begin{array}{lll}
\phi_1^{(1)}=m^{1\over2}(2\lambda)^{-{1\over 4}} &  & 
\phi_1^{(2)}=m\sqrt{2\over\lambda}\\
\phi_2^{(1)}={1\over 3}\phi_1^{(1)} &  & 
\phi_2^{(2)}={1\over 2}\phi_1^{(2)},
\end{array}
\eea
where superscripts $(1,2)$ refer to different potentials $U_{1,2}(\phi)$.
From these the height of the potential barrier can be calculated
$U(\phi_2^{(1,2)})={2m^3\over 27\sqrt{2\lambda}},{m^4\over 16\lambda}$.
The effect of changing $\lambda$ in $U_1$ and $U_2$ can be seen by
considering the associated Lagrangian (\ref{qlagrangian}) with a
potential of the form 
\be{apotential}
U(\phi)=b\phi^2(\phi^A-\sqrt{m^2\over 2 b})^2,
\ee
where $b$ is a constant. Clearly this potential is of the same form 
as $U_1$ and $U_2$ with the assumption that at $\omega=0$ there is a
degenerate minimum.
Under the scaling $\phi\rightarrow s\phi,\ b\rightarrow s^{-2A} b$
the Lagrangian transforms as $\cL\rightarrow s^2\cL$. Charge and energy
scale as $Q\rightarrow b^2Q\ E\rightarrow b^2 E$ while the radius
of the Q-ball is invariant as can be seen from the equation of motion
(\ref{eom}). From this we can see the effect of changing $\lambda_i
\rightarrow\lambda_i'$ ($i=1,2$) in 
$U_1$ and $U_2$ while keeping $m$ constant; charge and energy scale as
\bea{qandescaling}
Q & \rightarrow & ({\lambda_i\over\lambda_i'})^{1/A}Q\nonumber\\
E & \rightarrow & ({\lambda_i\over\lambda_i'})^{1/A}E,
\eea
where $A=2,1$ for $U_{1,2}(\phi)$, respectively.
The plotted figures therefore represent typical Q-ball solutions,
changing $\lambda$ simply scales the field $\phi$.

We will now consider some general properties of Q-ball solutions.
As $\omega$ increases from zero, a new global minimum $\phi_0$
with $U(\phi_0)<0$ develops. This does not, however, automatically
imply the existence of Q-balls as can be seen from the equation of motion
(\ref{eom}). If we consider the problem of solving the equation of motion
in terms of a mechanical analogy, where a Q-ball solution is analogous
to the motion of a particle in the potential $-U(\phi)$ with $r$ playing
the role of time,
we can easily see that the presence of the 'friction' term 
${2\over r}\dder{\phi}{r}$ requires that $\omega$ must be large enough 
to ensure that the initial 'potential energy' of the particle
is large enough for the particle to reach $\phi=0$. Put more
mathematically, integrating (\ref{eom}) with respect to $r$
and using the fact that $U_{\omega}(\phi(\infty))=0$ one gets
\be{inteom}
U_{\omega}(\phi(0))+2\int_0^\infty dr{1\over r}(\phi'(r))^2=0,
\ee
from which it is clear that $U_\w(\phi(0))$ must be large enough
to overcome the contribution from 'friction'.
As $\w$ tends to $m$ the potential barrier diminishes until at $\w=m$ 
it disappears. As the barrier disappears Q-ball solutions no longer
exist since the field now cannot stop at $\phi=0$ but will continue
rolling and acquire a negative value.

Numerically the solution can be found by taking into account the 
boundary conditions $\phi'(0)=0$ and $\phi(\infty)=0$.
The initial value of the field, $\phi(0)$, is then varied until a 
satisfactory solution is found. Depending on 
the value of $\w$ and the form of the potential, different types solutions 
can be found. The solutions vary from a thin-walled Q-ball to a thick-walled
type as $\w$ tends to $m$ in each case. This is as expected since as 
$\w$ starts to increase a global minimum develops and $\phi(0)$ lies
near the minimum leading to a thin-walled solution. As $\w$ increases
further
$\phi(0)$ moves further away from the global minimum leading to
thick-walled Q-balls.

\begin{figure}
\leavevmode
\centering
\vspace*{95mm}
\includegraphics{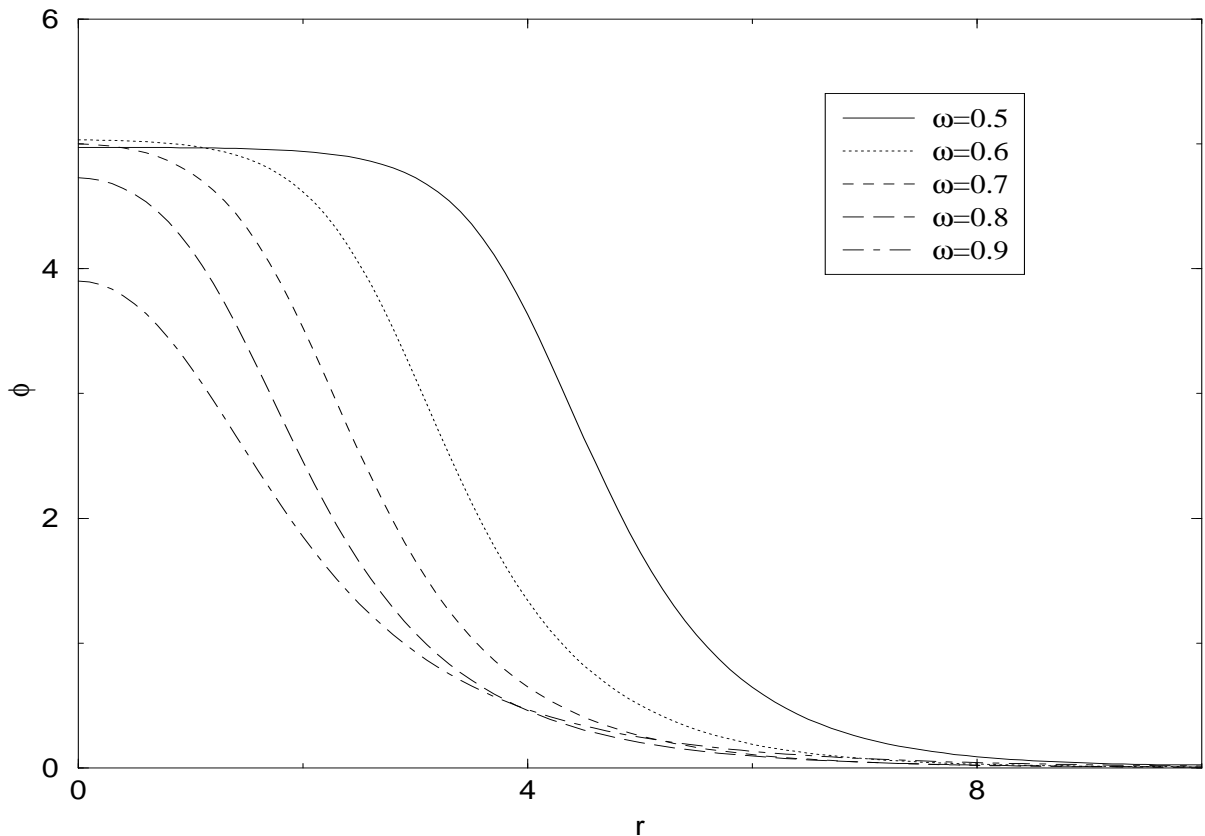}
\caption{$\phi$ as a function of $r$, $U=U_1(\phi)$ ($m_1=1$)}
\label{pot1profs}
\end{figure}

\begin{figure}
\leavevmode
\centering
\vspace*{95mm}
\includegraphics{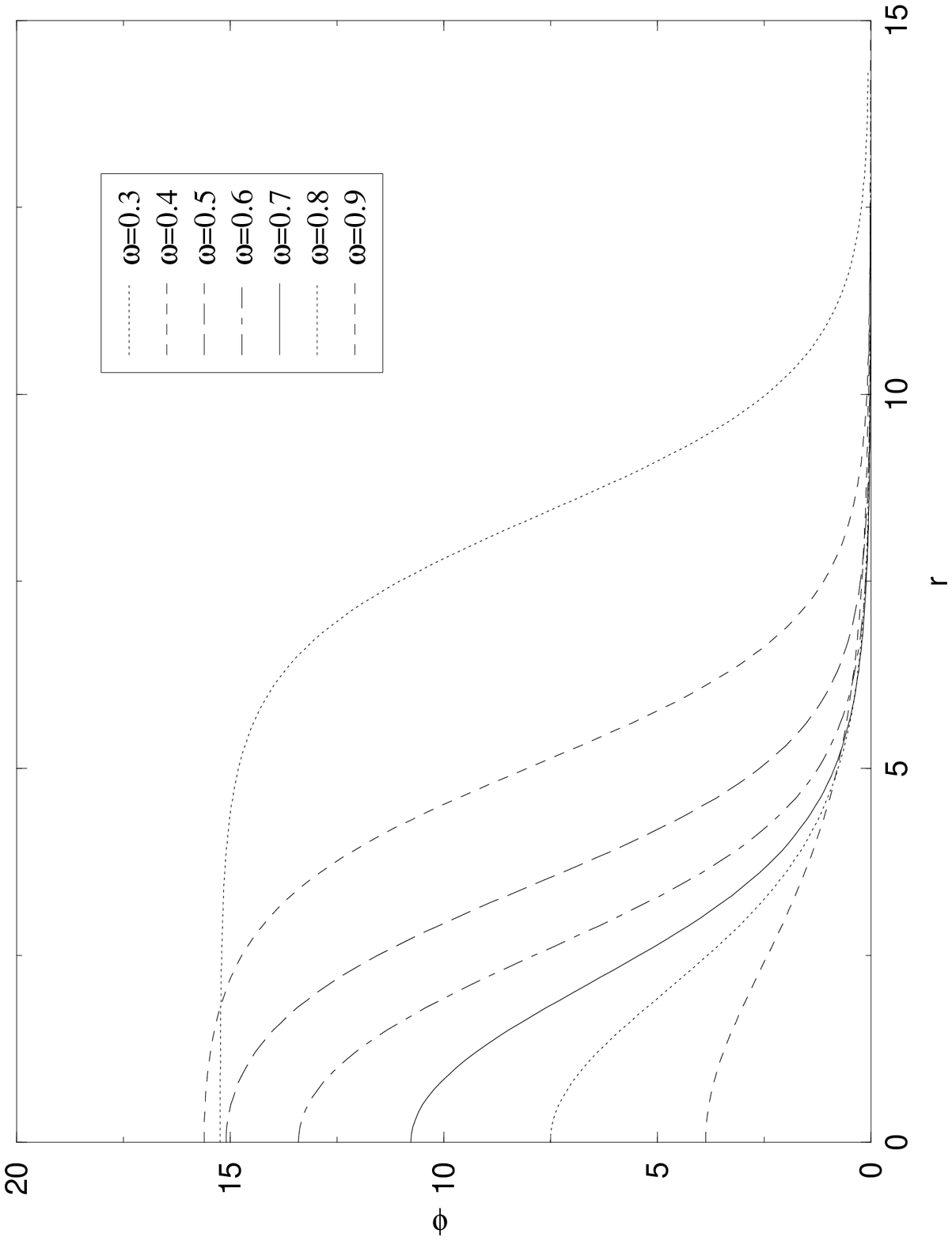}
\caption{$\phi$ as a function of $r$, $U=U_2(\phi)$ ($m_2=1$)}
\label{pot2profs}
\end{figure}

\begin{figure}
\leavevmode
\centering
\vspace*{95mm}
\includegraphics{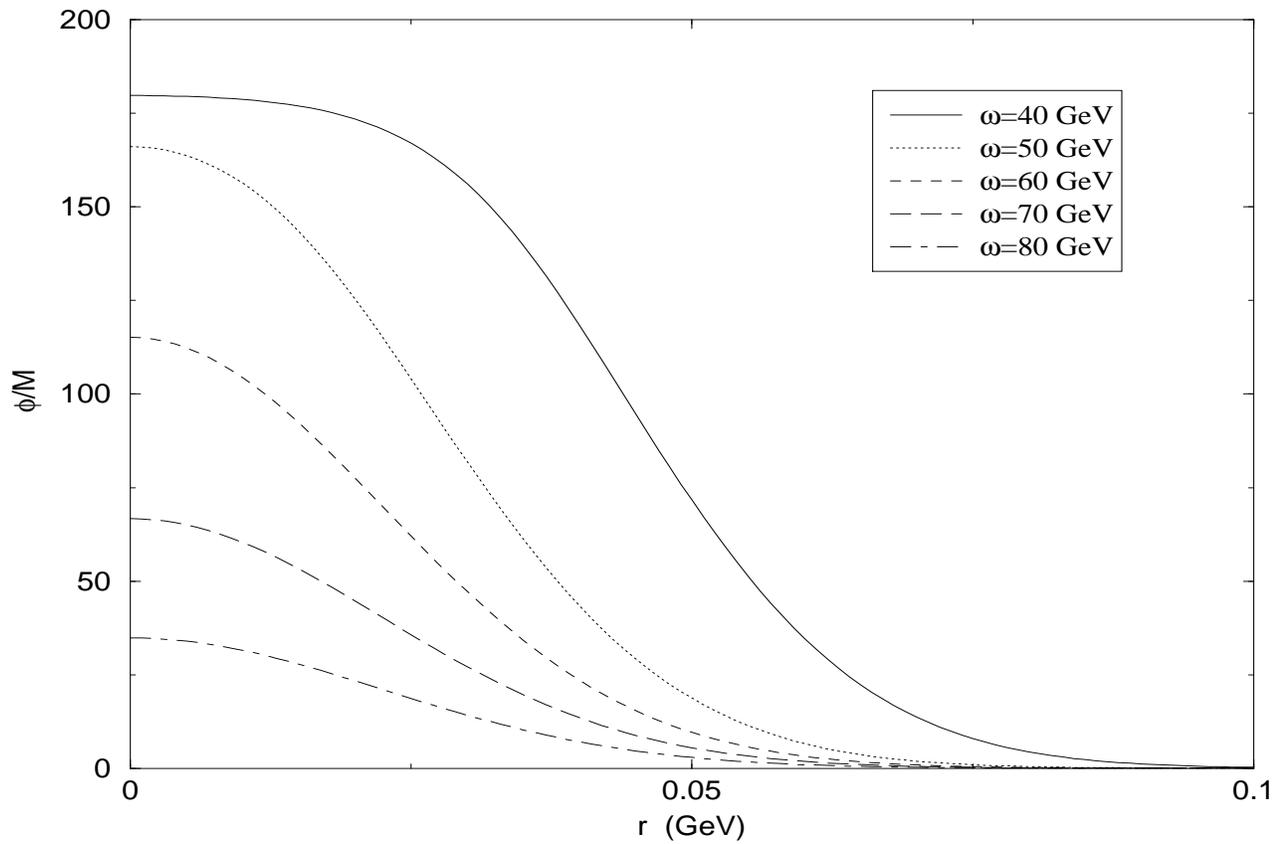}
\caption{$\phi$ as a function of $r$, $U=U_3(\phi)$}
\label{pot3profs}
\end{figure}

We have plotted the solutions for different values of $\w$ for each potential
for a typical choice of parameters in Figures \ref{pot1profs},
\ref{pot2profs} and \ref{pot3profs}. From
the Figures \ref{pot1profs} and \ref{pot2profs} the transformation from
a thin-walled ball to a thick-walled one can clearly be seen.

\begin{figure}
\leavevmode
\centering
\vspace*{95mm}
\includegraphics{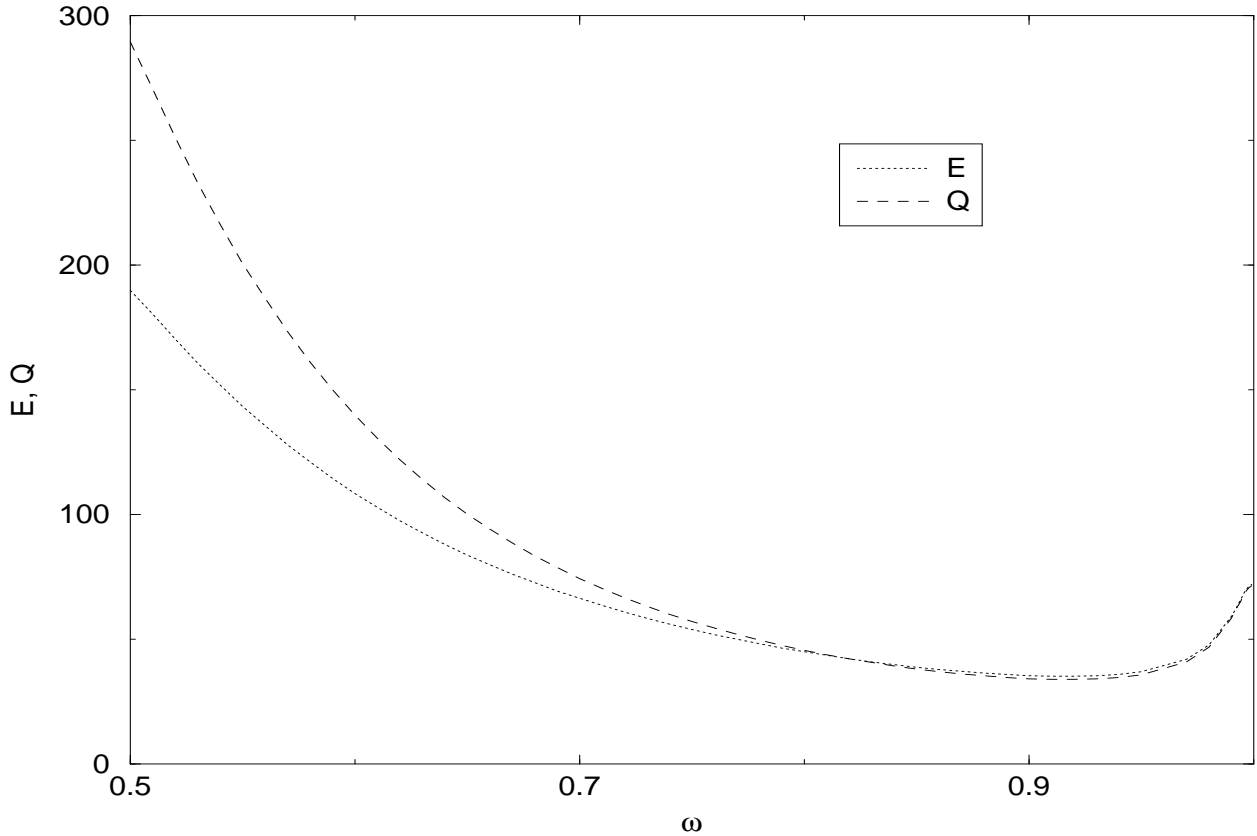}
\caption{Energy and charge versus $\w$, $U=U_1(\phi)$ ($m_1=1$)}
\label{pot1eandq}
\end{figure}

\begin{figure}
\leavevmode
\centering
\vspace*{95mm}
\includegraphics{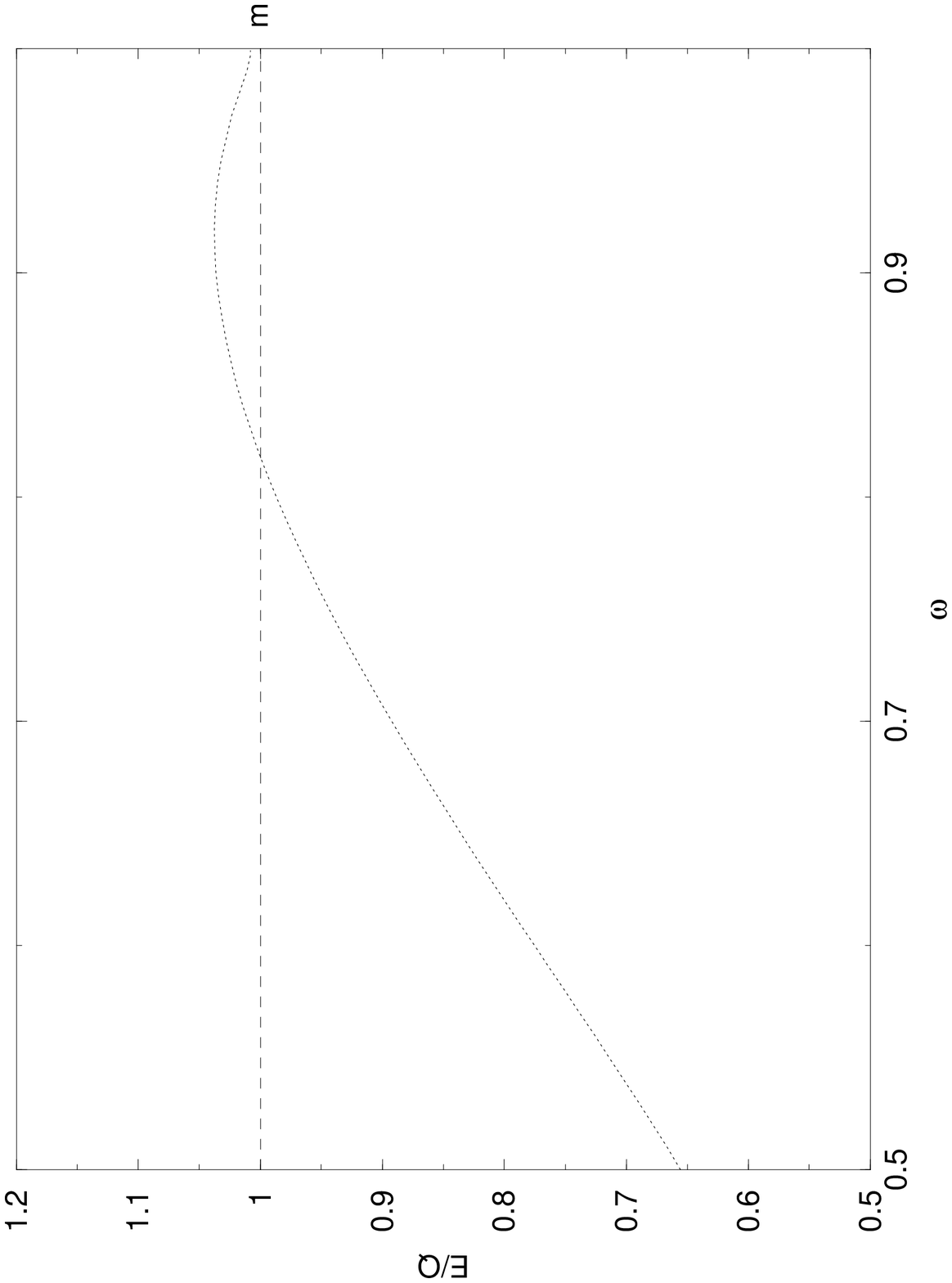}
\caption{Energy-charge ratio versus $\w$, $U=U_1(\phi)$ ($m_1=1$)}
\label{pot1eperq}
\end{figure}

\begin{figure}
\leavevmode
\centering
\vspace*{95mm}
\includegraphics{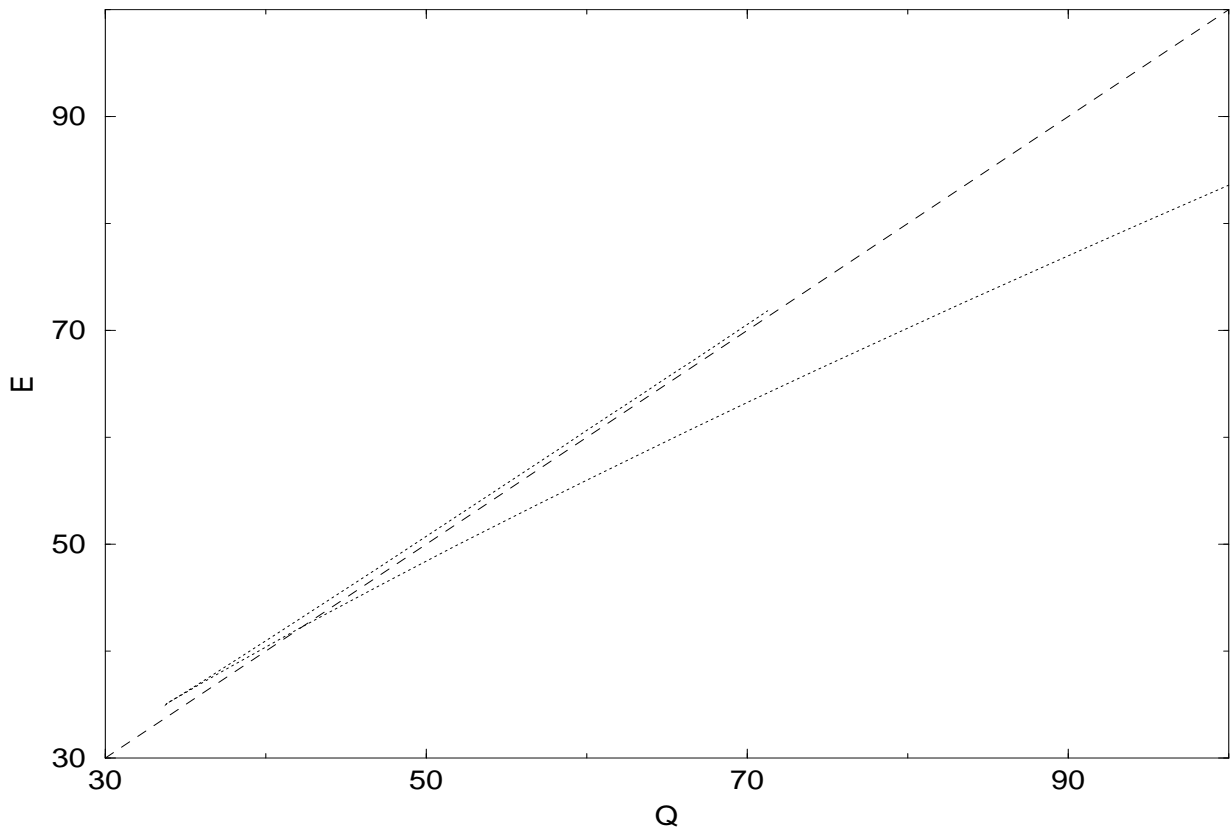}
\caption{Energy versus charge, $U=U_1(\phi)$ ($m_1=1$)}
\label{pot1evsq}
\end{figure}

\begin{figure}
\leavevmode
\centering
\vspace*{95mm}
\includegraphics{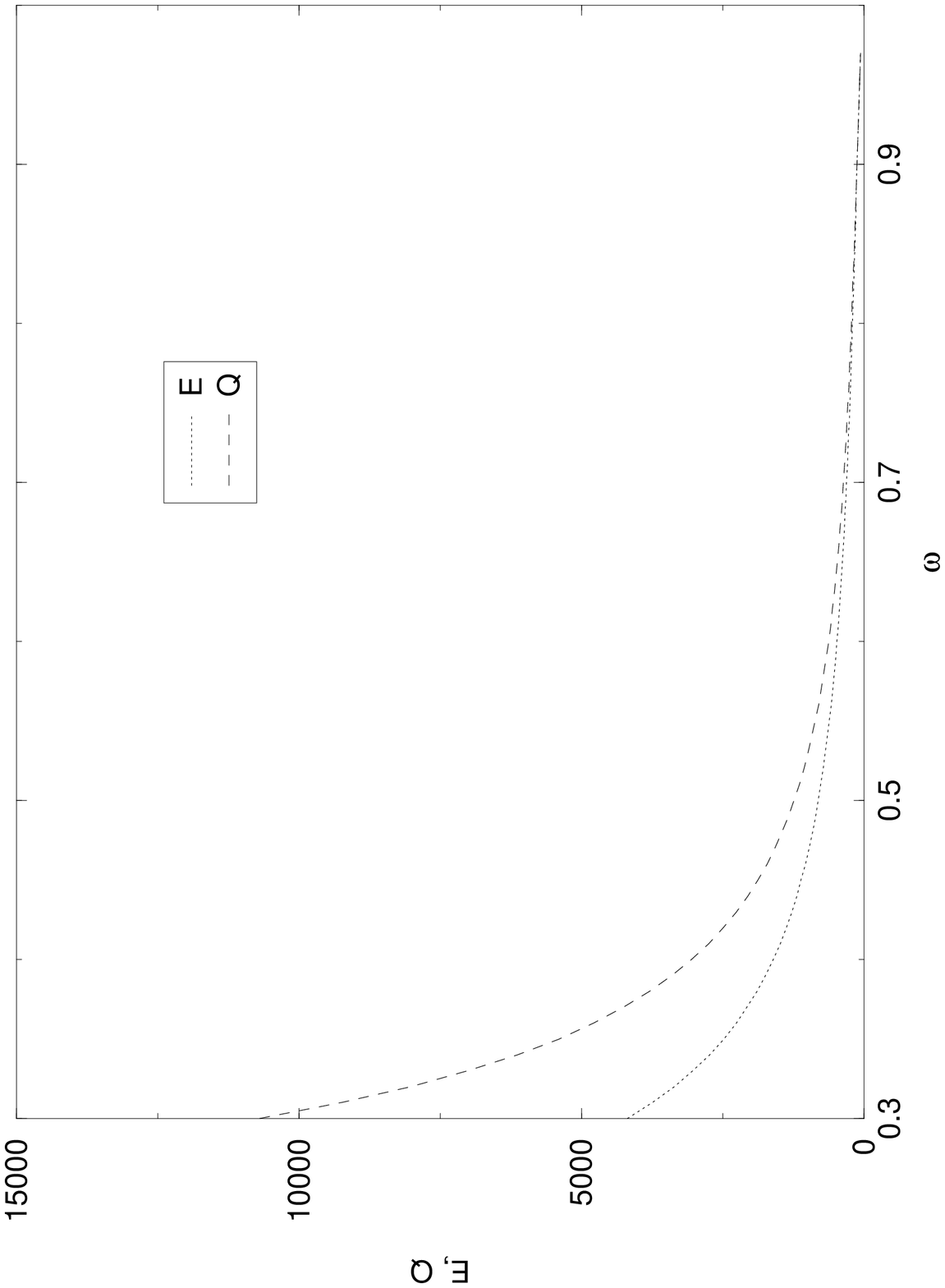}
\caption{Energy and charge versus $\w$, $U=U_2(\phi)$ ($m_2=1$) }
\label{pot2eandq}
\end{figure}

\begin{figure}
\leavevmode
\centering
\vspace*{95mm}
\includegraphics{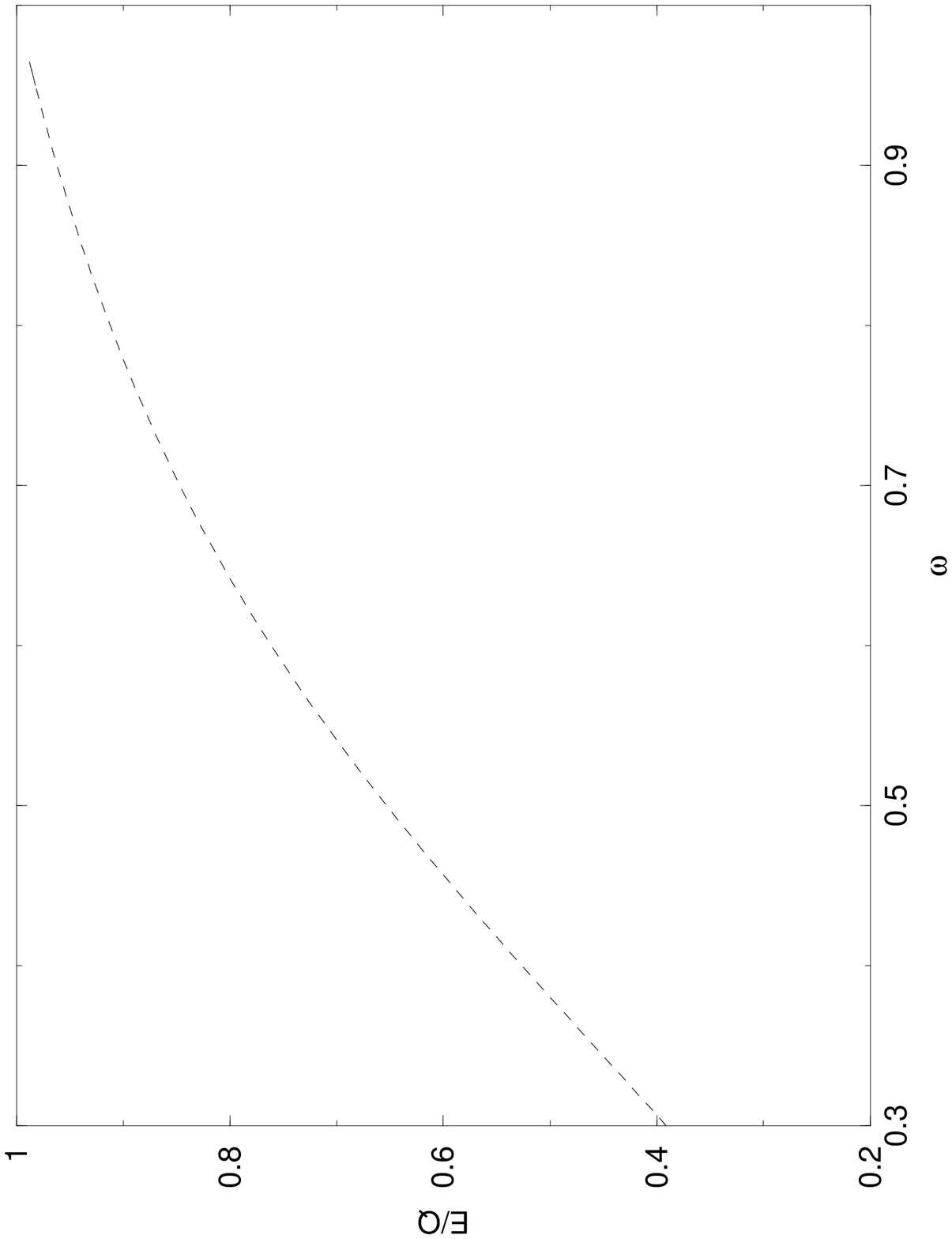}
\caption{Energy-charge ratio versus $\w$,  $U=U_2(\phi)$ ($m_2=1$)}
\label{pot2eperq}
\end{figure}

\begin{figure}
\leavevmode
\centering
\vspace*{95mm}
\includegraphics{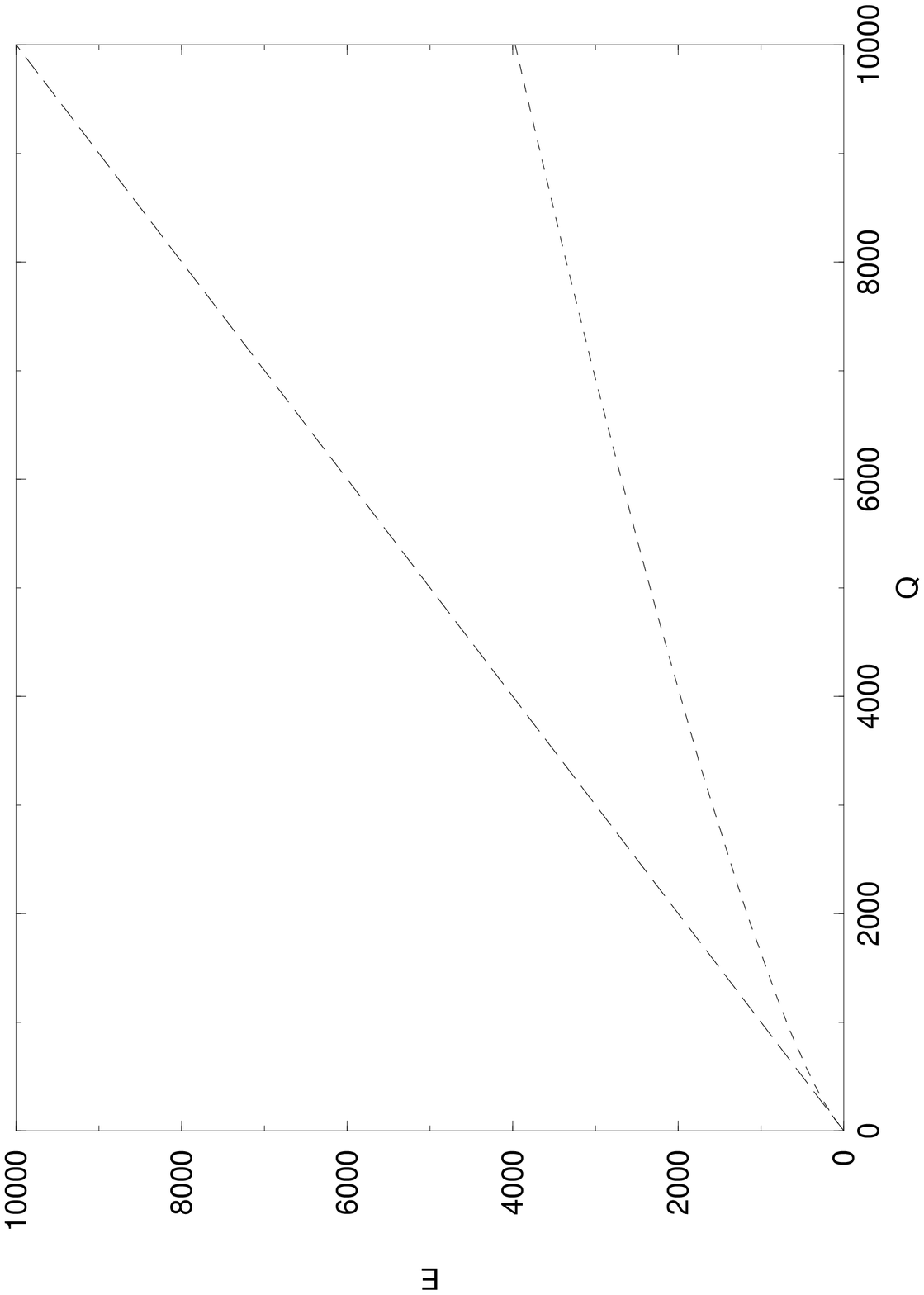}
\caption{Energy versus charge, $U=U_2(\phi)$ ($m_2=1$)}
\label{pot2evsq}
\end{figure}

\begin{figure}
\leavevmode
\centering
\vspace*{95mm}
\includegraphics{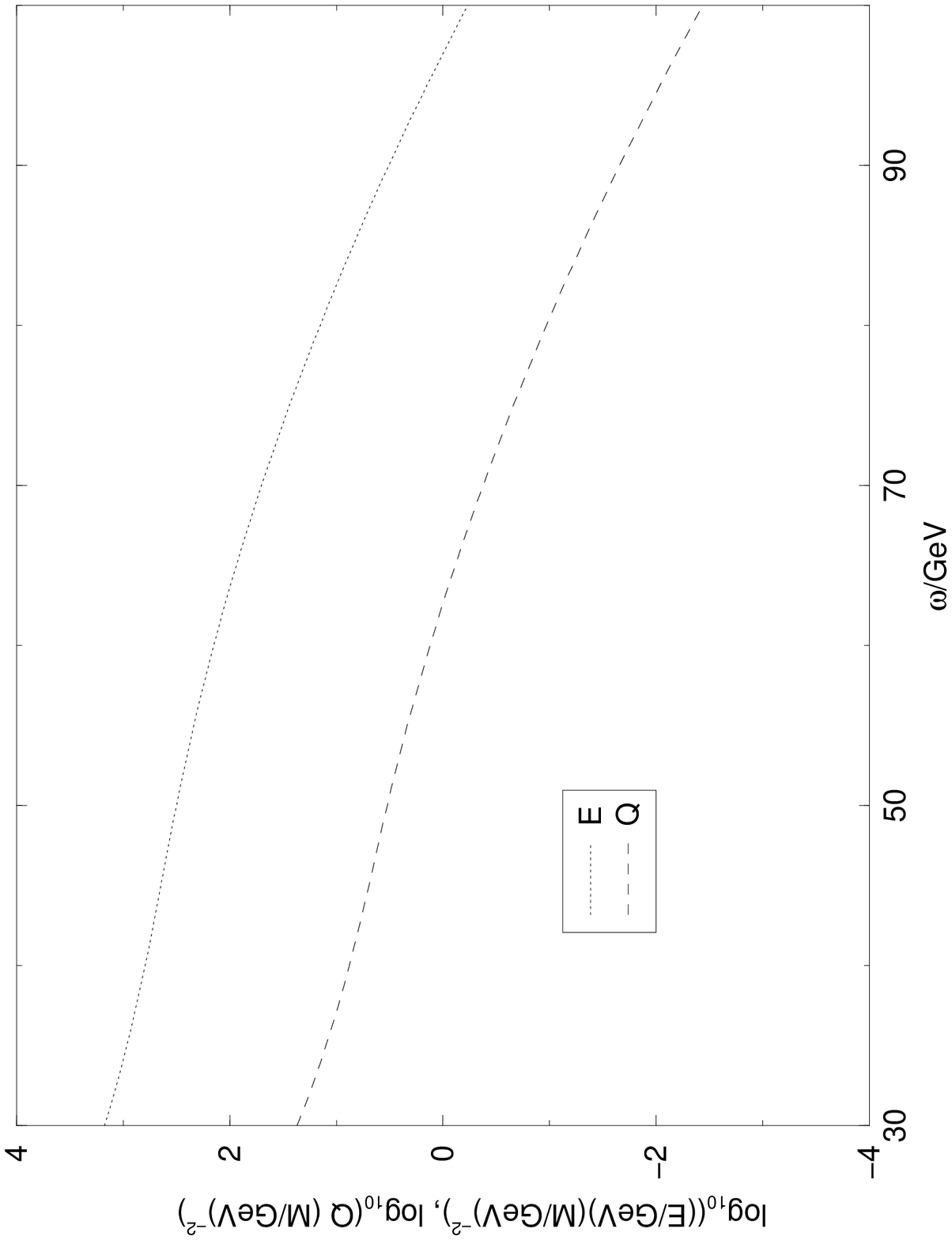}
\caption{Energy and charge versus $\w$, $U=U_3(\phi)$}
\label{pot3eandq}
\end{figure}

\begin{figure}
\leavevmode
\centering
\vspace*{95mm}
\includegraphics{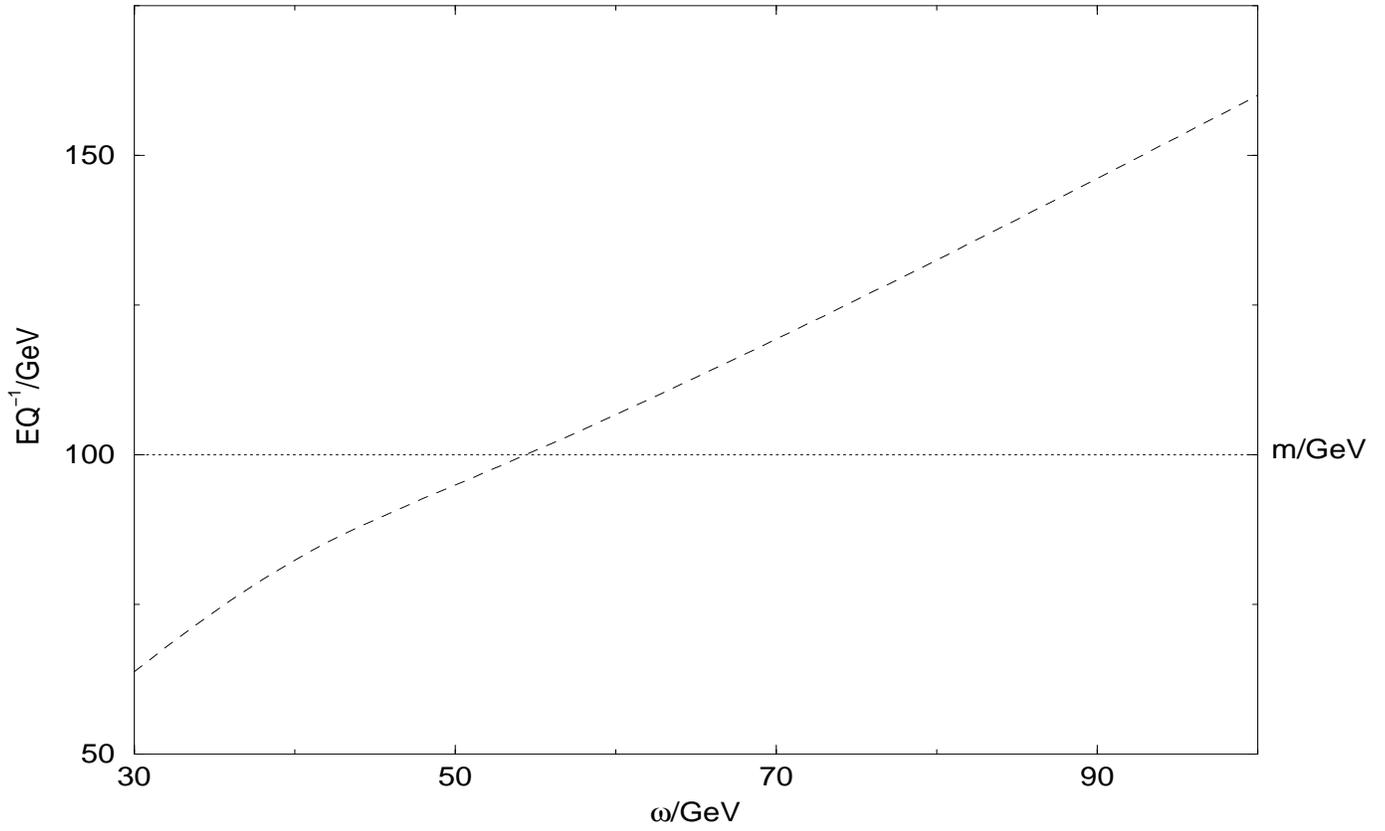}
\caption{Energy-charge ratio versus $\w$,  $U=U_3(\phi)$}
\label{pot3eperq}
\end{figure}

\begin{figure}
\leavevmode
\centering
\vspace*{95mm}
\includegraphics{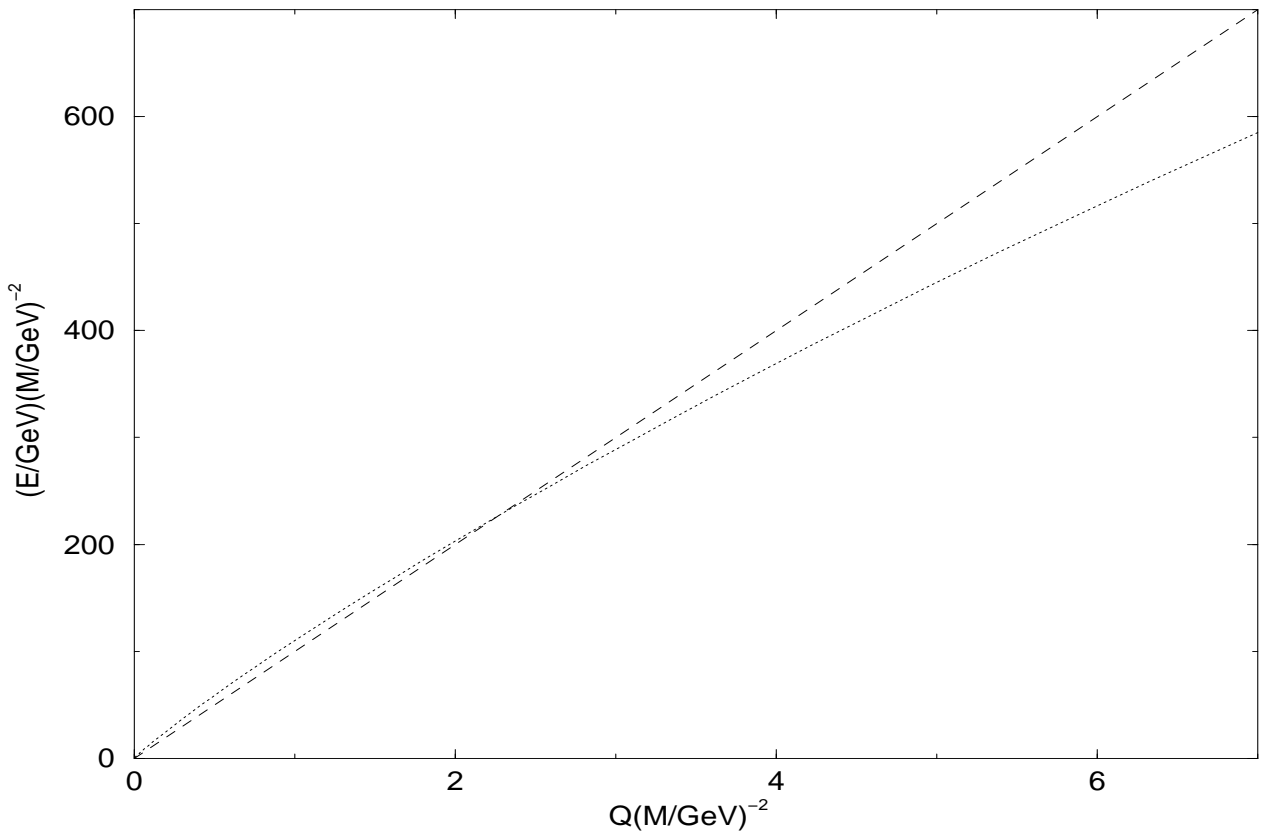}
\caption{Energy versus charge, $U=U_3(\phi)$}
\label{pot3evsq}
\end{figure}

The charge and energy of a Q-ball are given by (\ref{Q3},\ref{E3}). We have
plotted the charge and energy versus $\w$, $E$ as a function
of $Q$ and $E/Q$ versus $\w$
for the different types of potentials in Figures 
\ref{pot1eandq}-\ref{pot3evsq}.
In the figures where $E$ is plotted versus $Q$, Figures 
\ref{pot1evsq}, \ref{pot2evsq} and \ref{pot3evsq}, a straight line $E=mQ$
indicating stability is drawn, Q-balls are stable (with respect to
decays into $\phi$ scalars) under this line.
In all the cases there is a different behaviour as $\w$ tends to $m$.

For $U_1(\phi)$ there exists a region of small $\w$ where stable
thin-walled Q-balls exist. As
$\w$ increases charge decreases and Q-balls become thick-walled.
As charge continues to decrease Q-balls first become unstable.
Charge and energy then reach a minimum and start to grow again, while
still in the region of instability. As $\w$ tends to one, $E$ tends
asymptotically to $mQ$. This behaviour is also clear from
the $E/Q$ versus $\w$ figure, Fig. \ref{pot1eperq}, where a horizontal line at
$E/Q=m$ 
represents the stability line.

Even though the Q-ball profiles in the second case, $U=U_2(\phi)$, look 
quite similar to those of the previous case, the charge and energy have
a different behaviour as $\w$ increases. The profile still mutates from
a thin-walled ball to a thick-walled one like before but now the energy
versus charge plot, Fig. \ref{pot2evsq}, shows a new type of behaviour. 
Clearly $E/Q<m$ for all $\w$
indicating that Q-balls are stable in this potential. The difference in 
Q-ball stability at the thick-wall limit in the two potentials 
$U_1(\phi)$ and $U_2(\phi)$ is in accordance
with the previous analytical reasoning (\ref{qconds}). 

The third case is, again, different from the two previous ones.
The stability curve, Fig. \ref{pot3evsq}, is now such that as 
$\w$ increases, Q-balls become
unstable like in the first case but here the charge and energy
simply decrease with increasing $\w$ for the whole range of parameter
values.

\clearpage

The numerical studies of these three types of potentials show different
behaviour of charge and energy with changing $\w$ for each type. As
analytical results in the previous section showed the behaviour in the
thick-wall limit depends critically on the power of the next to
leading term. The logarithmic potential also shows a different
kind of behaviour in the thick-wall limit. The question of stability
of Q-balls in the thick wall-limit is hence strongly dependent 
on the exact form of the potential.

\section{Q-ball evaporation}
In realistic theories where the $\phi$-field couples to other fields
the simple criterion for absolute stability $E<mQ$ no longer holds 
(it is still obviously true for $\phi$ decays).
Here we discuss a decay mode to massless fermions.
The evaporation of Q-balls to neutrinos was discussed in detail in
\cite{cohen}.
It was shown that for a large Q-ball with a step-function boundary 
there exists an upper bound of the evaporation rate per unit area.

We follow closely the approach described in \cite{cohen}.
This is a leading order semi classical approximation where 
massless fermions are considered
in the presence of a classical field, the Q-ball background.
Since the Dirac sea is filled inside the Q-ball, 
evaporation can only progress through the Q-ball surface and is hence
proportional to the area of the Q-ball. 
Evaporation progresses through pair-production at the Q-ball surface
so that each pair produced decreases the charge of the Q-ball by two
(e.g. a lepton number carrying ball could evaporate into neutrinos
so that $\Delta L=2$ for each pair of neutrinos produced).

The Lagrangian density is given by
\be{ld}
\cL=\partial_\mu
\phi^*\partial^\mu\phi-U(\abs{\phi})+\psi^\dagger(i\partial_0+i\sigma\cdot
\nabla)\psi-ig\phi\psi^\dagger\sigma_2\psi^*+ig\phi^*\psi^T\sigma_2\psi,
\ee
where $\psi$ is a two component Weyl spinor corresponding to the fermion
field, $\phi$ is the Q-ball scalar field and $\sigma$ are the Pauli matrices.
The equations of motion are
\bea{eqmotion}
(i\partial_0+i\sigma\cdot\nabla)\psi-g\phi\chi & = & 0\non\\
(i\partial_0-i\sigma\cdot\nabla)\chi-g\phi^*\psi & = & 0,
\eea
where $\chi=i\sigma_2\psi^*$ is a Weyl spinor. As in \cite{cohen}
we choose normal modes such that $\psi\sim e^{[-({\w\over 2}+\w')it]}$
and $\chi\sim e^{[-(-{\w\over 2}+\w' )it]}$. With this choice
the field equations become
\bea{eqmotion2}
({\w\over 2}+\w'+i\sigma\cdot\nabla)\psi-g\phi\chi & = & 0\non\\
(-{\w\over 2}+\w'-i\sigma\cdot\nabla)\chi-g\phi\psi & = & 0.
\eea
The  eigensolutions to these equations can be found using spherical 
spinors. We make the ansatz 
\bea{ansatz}
\psi & = & g_1(r)\Omega_{jlm}+if_1(r)\Omega_{jl'm}\non\\
\chi & = & g_2(r)\Omega_{jlm}+if_2(r)\Omega_{jl'm},
\eea
where $\Omega_{jlm},\ \Omega_{jl'm}$ are spherical spinors \cite{greiner},
\bea{spherics} 
\Omega_{j,\underbrace{\scriptstyle{j-1/2}}_l,m} & = & {\sqrt{j+m\over 2j} Y_{l,m-1/2}\choose
\sqrt{j-m\over 2j} Y_{l,m+1/2}}\non\\
\Omega_{j,\underbrace{\scriptstyle{j+1/2}}_l,m} & = & {-\sqrt{j-m+1\over 2j} Y_{l,m-1/2}\choose
\sqrt{j+m+1\over 2j+2} Y_{l,m+1/2}},
\eea
$Y_{lm}$ are spherical harmonics,
$j$ and $m$ are the angular momentum
quantum numbers and $l'=2j-l$ and $j=l\pm {1\over 2}$.

Substituting the ansatz into the field equations (\ref{eqmotion2})
the radial equations are easily found (we set $g=1$):
\bea{radeqs}
({\w\over 2}+\w')g_1(r)+f_1'(r)-f_1(r){{3\over 2}+j\over r} & = & 
\phi(r)g_2(r)\non\\
({\w\over 2}+\w')f_1(r)-g_1'(r)-g_1(r){{1\over 2}-j\over r} & = & 
\phi(r)f_2(r)\\
(-{\w\over 2}+\w')g_2(r)+f_2'(r)+f_2(r){{3\over 2}+j\over r} & = & 
\phi(r)g_1(r)\non\\
(-{\w\over 2}+\w')f_2(r)+g_2'(r)+g_2(r){{1\over 2}-j\over r} & = & 
\phi(r)f_1(r)\non.
\eea
We must now solve these equations with appropriate boundary conditions.
Near the origin $r=0$ we can approximate the Q-ball potential with
a constant, $\phi(r)=\phi_0$. In a suitable basis the radial equations
decouple and an analytical solution can be found:
\bea{radeqs0}
g_1^0(r) & = & A_1 (\eta+\alpha) \eta_1 j_{l_-}(\eta_1r)+A_2(\eta-\alpha)
\eta_2 j_{l_-}(\eta_2r)\non\\
f_1^0(r) & = & A_1 (\w+\alpha) \eta_1 j_{l_+}(\eta_1r)+A_2(\w-\alpha)
j_{l_+}(\eta_2r)\\
g_2^0(r) & = & A_1 g\phi_0 \eta_1j_{l_-}(\eta_1r)+A_2 g\phi_0 \eta_2 
j_{l_-}(\eta_2r)\non\\
f_2^0(r) & = & A_1 g\phi_0 \eta_1j_{l_+}(\eta_1r)+A_2g\phi_0\eta_2
j_{l_+}(\eta_2r)\non,
\eea
where $l_{\pm}=j\pm{1\over 2}$, $\alpha=\sqrt{\w'^2-g^2\phi_0^2}$
and $\eta_{1,2}={\w'\over 2}\pm\alpha$.
Here the divergent part (at the origin) of the general solutions has been 
omitted to keep the wave functions normalizable at $r=0$.
Far away from the Q-ball surface $\phi=0$ and the equations of motion
decouple into two pairs. An analytical solution can again be found:
\bea{radeqsinf}
g_i^\infty(r) & = & A_i k_i h^{(1)}_{l_-}(k_i r)+B_i k_i h^{(2)}_{l_-}(k_i r)\non\\
f_i^\infty(r) & = & A_i k_i h^{(1)}_{l_+}(k_i r)+B_i k_i h^{(2)}_{l_+}(k_i r),
\eea
where $i=1,2$, $h^{(i)}$ are the spherical Hankel functions and
$k_{1,2}={\w\over 2}\pm\w'$.
From these we can identify the in- and out-moving waves since 
$h^{(1)}\sim e^{ir}$
and $h^{(2)}\sim e^{-ir}$. Solving for the coefficients $A_i$ and $B_i$ we get
\bea{coeff}
A_i & = & {1\over k_i} {g_i^\infty(r) h^{(2)}_{l_+}(k_i r)-f_i^\infty(r) 
h^{(2)}_{l_-}(k_i r)\over
h^{(1)}_{l_-}(k_i r)h^{(2)}_{l_+}(k_ir)-h^{(1)}_{l_+}(k_ir)h^{(2)}_{l_-} (k_ir)}\non\\ 
B_i & = & {1\over k_i}{g_i^\infty(r) h^{(1)}_{l_+}(k_i r)-f_i^\infty(r) h^{(1)}_{l_-}(k_i r)\over
h^{(2)}_{l_-}(k_i r)h^{(1)}_{l_+}(k_ir)-h^{(2)}_{l_+}(k_ir)h^{(1)}_{l_-}(k_ir)}.
\eea

The evaporation rate can be calculated assuming that there is no
incoming $\chi$ wave since then all the reflected $\chi$ flux must
have been transmutated. From the transmutation coefficient $T$
the evaporation rate can be calculated \cite{cohen}:
\be{evaprate}
{dQ\over dt}=\sum_j\int_0^{\w/2}{dk\over 2\pi}(2j+1)|T(k,j)|^2.
\ee
Here it is worth noting another result obtained in \cite{cohen},
\be{transrelat}
|T(k_+,j)|=|T(k_-,j)|\leq 1.
\ee

Assuming that the potential $U(\phi)$ scales as $U(s\phi)=s^2U(\phi)$,
we note that under the transformation 
$\phi\rightarrow s\phi$, $\psi\rightarrow s\psi$,
$g\rightarrow s^{-1}g$
the Lagrangian (\ref{ld}) transforms as $\cL\rightarrow s^2\cL$
so that the evaporation rate is invariant.
Assuming as before a potential of the form (\ref{apotential})
we note that the evaporation rate is invariant if $b$ scales
as $b\rightarrow s^{-2A}b$. 

In \cite{cohen} the step-function boundary was considered in the large R 
limit. The calculation for any R is easily done by matching the in and 
out solutions (\ref{radeqs0},\ref{radeqsinf}). Assuming that there
is no incoming $\chi$ flux, all the reflected $\chi$ waves must have then 
been transmutated. Identifying the coefficients $A_i,\ B_i$ with the incoming, 
reflected and transmutated waves,
the transmutation rate can be calculated in a straightforward way.
The resulting expression is quite lengthy and is not presented here.

\begin{figure}
\leavevmode
\centering
\vspace*{95mm}
\includegraphics{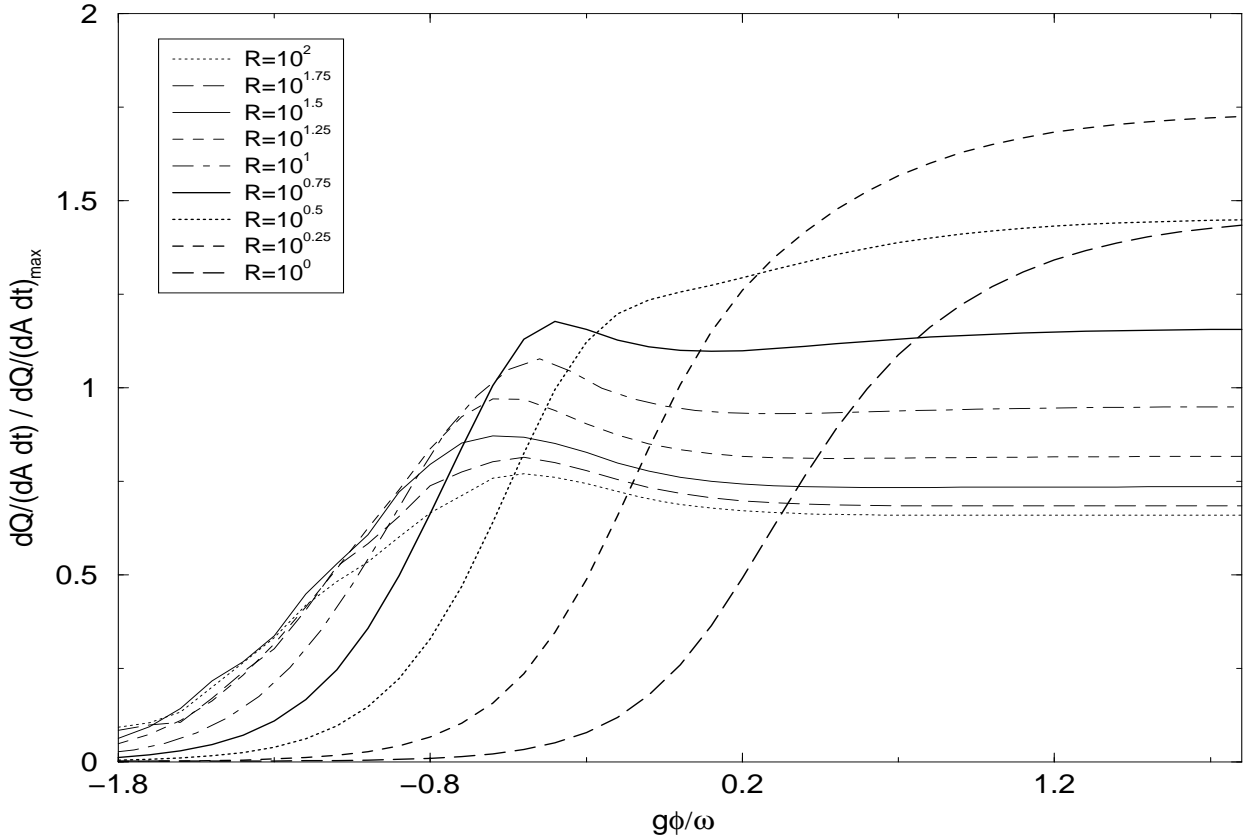}
\caption{Evaporation rates for a step-function boundary}
\label{steprate}
\end{figure}

The maximum evaporation rate per unit area in the $R\rightarrow\infty$ limit 
was calculated in \cite{cohen} by trading the angular momentum variable
$j$ for a linear momentum variable and using relation (\ref{transrelat}).
The maximum evaporation rate in this limit is given by
\be{maxrate}
({dQ\over dA dt})_{\rm{max}}={\w^3\over 192\pi^2}.
\ee
We have plotted the evaporation rates per unit area as a fraction
of $({dQ\over dA dt})_{\rm{max}}$ for a number of $R$'s in Fig.
\ref{steprate}.

Obviously the evaporation rate is strongly dependent
on R and it approaches the limiting profile given in \cite{cohen}.
From the Fig. \ref{steprate} it can be seen
that the evaporation rate per unit area first increases with $R$
for all $g\phi_0/\w$ until it reaches a maximum. Then the maximum value
slowly starts
to decrease and shift towards smaller $g\phi_0/\w$ while a ``bump''
develops around $\log(g\phi_0/\w)\sim -0.5$. The different types of
behaviour at small and large $g\phi_0/\w$ is due to the
presence of the $\exp(i\sqrt{\w'^2-g^2\phi_0^2}R)$-term in the matching 
process of the wave functions in- and outside the Q-ball. 
When $\w'>g\phi_0$ the
wave-function consists of oscillatory functions compared to
$\w'<g\phi_0$ when exponentially growing and decaying parts are present.
As $R$ increases this effect becomes more pronounced and the evaporation
rate tends to a constant more quickly as $g\phi_0/\w$ increase. 
The slightly rough
features at small $g\phi_0/\w$ are also due to the strong oscillatory
$R$ dependence.

In a realistic case a step-function is not always a good approximation.
It approximates a thin-walled Q-ball well but for a thick-walled
Q-ball its accuracy may be questioned. The profile a thick-walled
Q-ball is significantly different from a step-function and furthermore
its radius is not well defined. To model a thick-walled Q-ball with 
a step-function Q-ball one could for example choose the charges to be
equal. This would still leave the radius and the value of the field
inside the ball arbitrary. One should then define the Q-ball boundary to
obtain a radius and hence obtain $\phi_0$. 
The accuracy of such an approximation can be questioned especially
in the case of very thick-walled Q-balls.

We have examined the evaporation rates numerically for realistic Q-balls 
profiles (in addition to checking that the purely numerical calculation
recovers the results for a step-function boundary).
This is done by solving the equations (\ref{radeqs})
numerically for a given profile. The profiles have been calculated
from the different types of potentials, $U_i(\phi),\ i=1,2,3$.

The numerical solution is found using the fact that asymptotically
the Q-ball profile at the origin is flat and can be approximated
with a constant. The analytical solution is then given by (\ref{radeqs0}).
Far away from the Q-ball surface the potential is zero and the
solution there is given by (\ref{radeqsinf}). The problem then reduces
to finding the right initial conditions to obtain a solution
from which we can deduce the transmutation coefficient \ie 
we must choose $A_1$ and $A_2$ in (\ref{radeqs0}) such that
far away from the Q-ball there is no incoming $\chi$-wave.
The coefficients for the incoming and reflected $\psi$-wave
and for the transmutated $\chi$-wave can then be found by 
using (\ref{coeff}).

\begin{figure}
\leavevmode
\centering
\vspace*{95mm}
\includegraphics{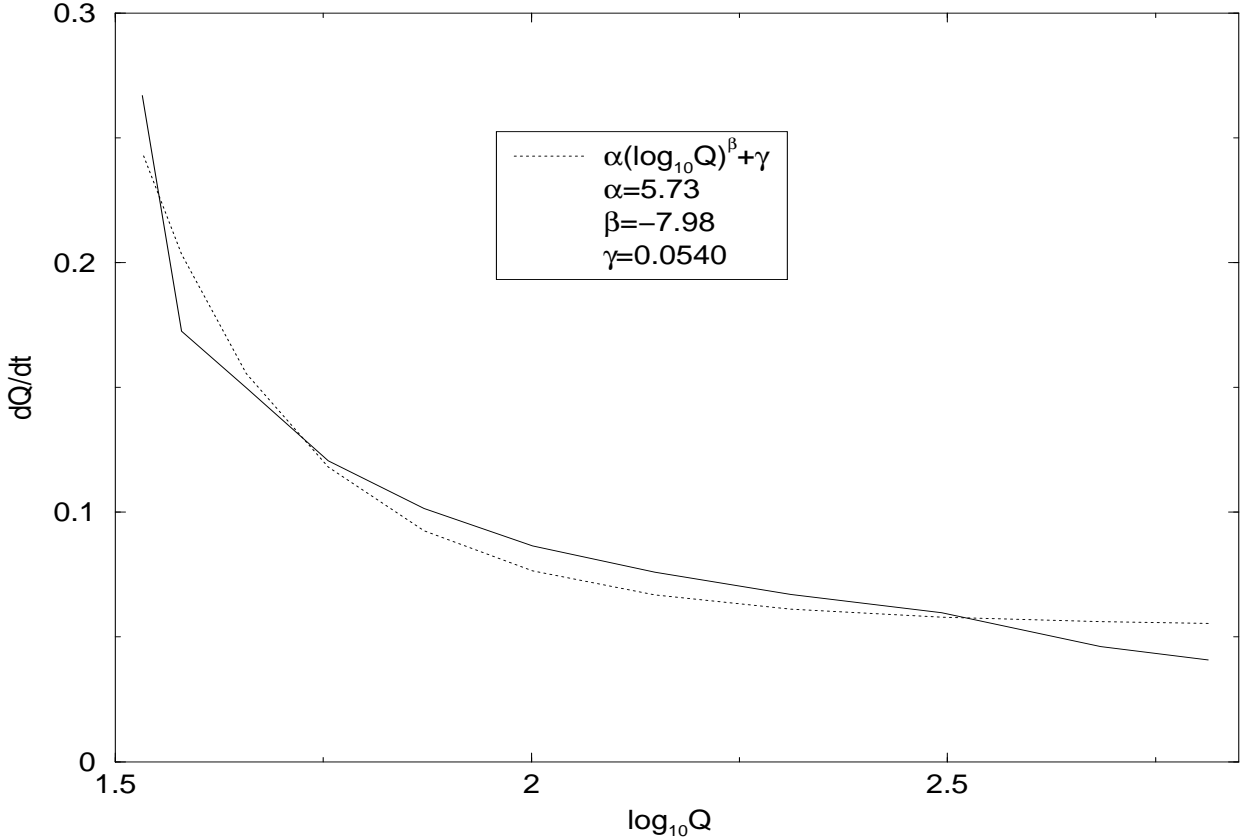} 
\caption{Evaporation rate ${dQ\over dt}$ versus $\log Q$, 
$U=U_1(\phi)$ ($m_1=1$)}
\label{pot1evap}
\end{figure}

\begin{figure}
\leavevmode
\centering
\vspace*{95mm}
\includegraphics{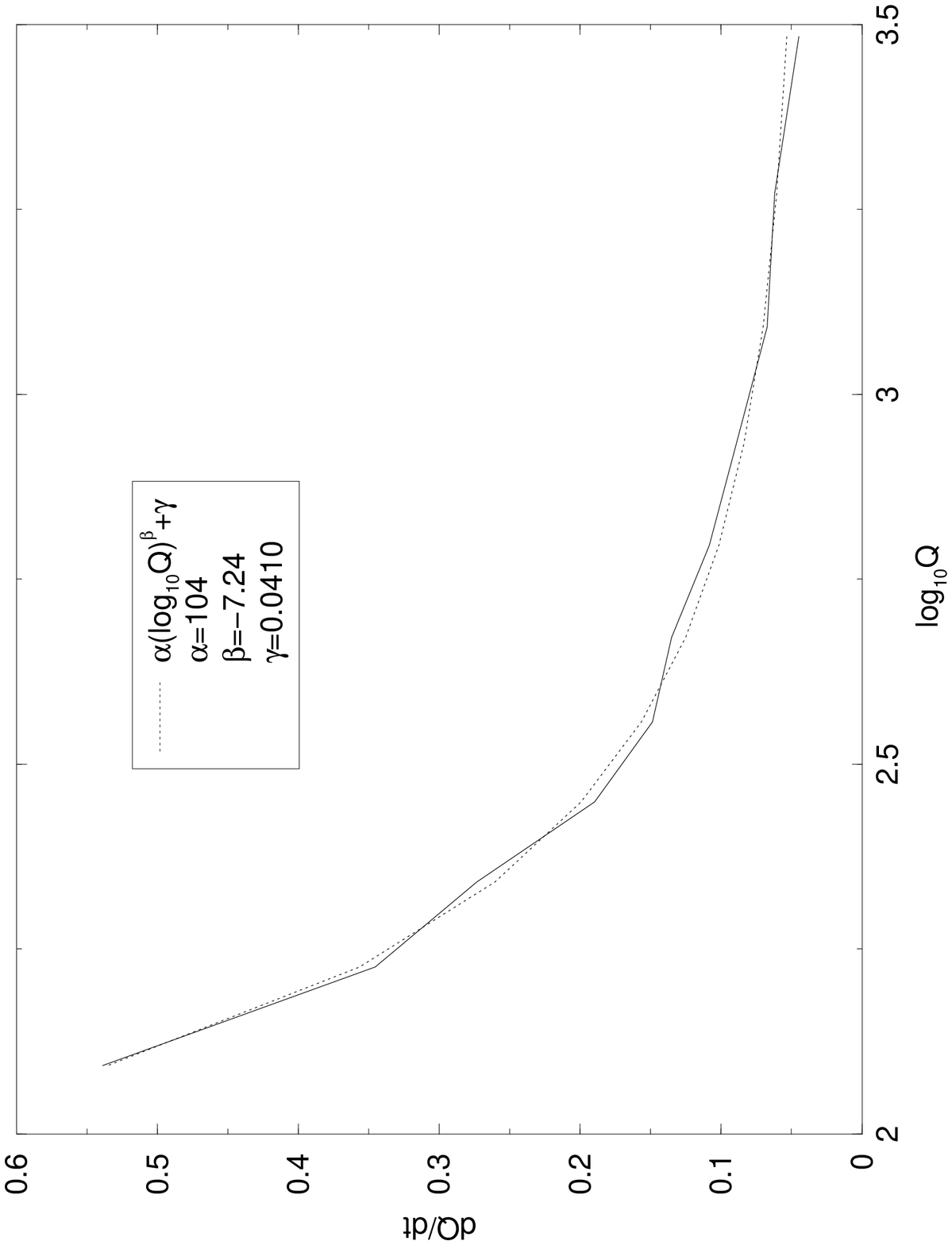}
\caption{Evaporation rate ${dQ\over dt}$ versus $\log Q$, 
$U=U_2(\phi)$ ($m_1=1$)}
\label{pot2evap}
\end{figure}

\begin{figure}
\leavevmode
\centering
\vspace*{95mm}
\includegraphics{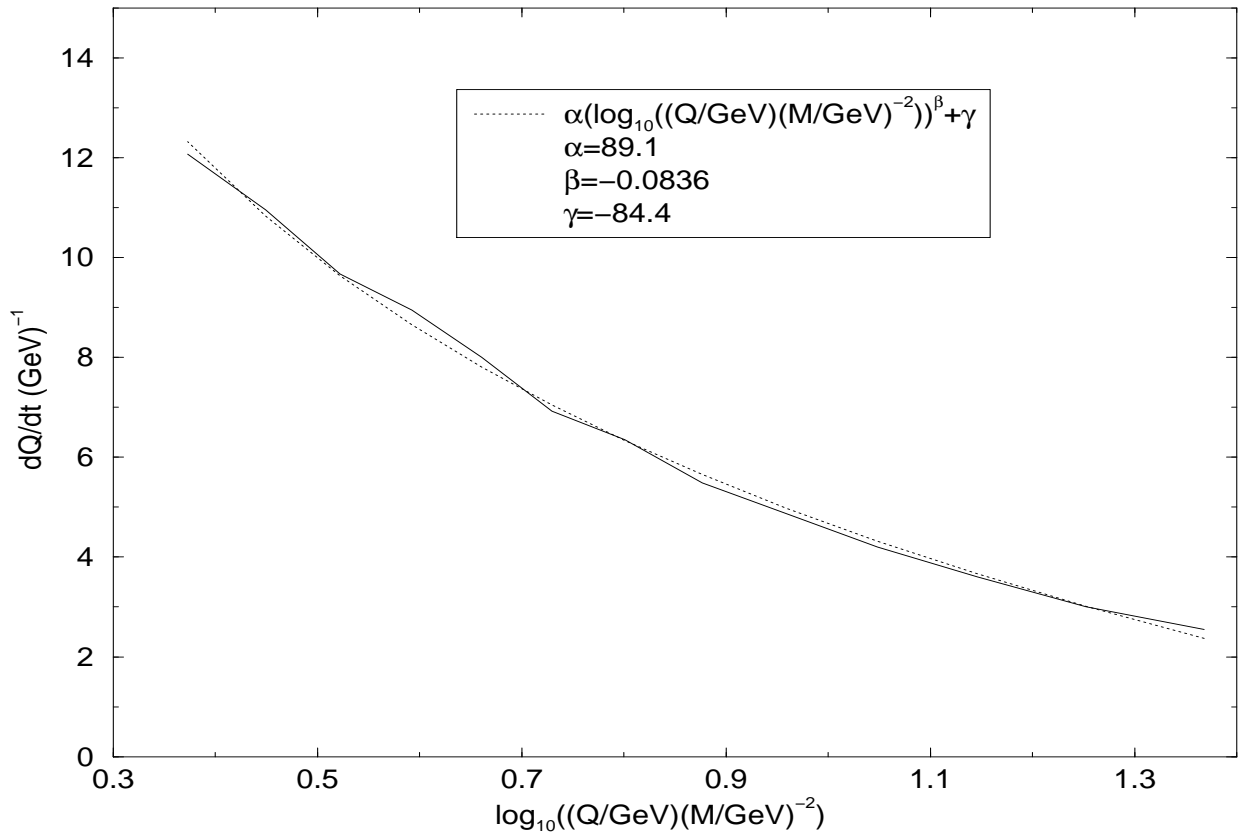}
\caption{Evaporation rate ${dQ\over dt}$ versus $\log Q$, 
$U=U_3(\phi)$}
\label{pot3evap}
\end{figure}

We have calculated the evaporation rate (\ref{evaprate})
for a number of $\w$:s in the three different potentials.
These are presented in Figures \ref{pot1evap}, \ref{pot2evap} and 
\ref{pot3evap} as a function of charge. Note that here the evaporation
rate is not given as a fraction of the maximum rate and is the total
evaporation rate (instead of the evaporation rate per unit area).
We have only plotted the evaporation rates for a parameter
range where Q-balls are stable with respect to decay into
$\phi$-scalars.

From the figures we see that at large Q the evaporation rate decreases
with increasing Q. This holds for the whole of the studied parameter
range for all the potentials so that a Q-ball
will evaporate faster as it decreases in size. 
The calculation is quite time consuming and requires high 
numerical accuracy in choosing $A_1$ and $A_2$ correctly.
This fact is evident in the slight non-smoothness present in the
figures due to numerical inaccuracies during the computation.

We have fitted a curve ${dQ\over dt}=\alpha(\log Q)^\beta+\gamma$
through the points in each case. These fits are presented in the 
corresponding figures.

It is cosmologically interesting to estimate the evaporation times,
especially for the realistic potential $U_3(\phi)$. Let us consider
a typical Q-ball in the logarithmic potential that is stable with
respect to scalar decays. 
For a Q-ball to survive past the electroweak (EW) phase transition
we can estimate that its lifetime must be $\sim 10^{-12}\ \textrm{s}\sim
10^{12}\ \textrm{(GeV)}^{-1}$. 
From the Fig. \ref{pot3evap} we can estimate that the evaporation
rate is $\sim 10\ \textrm{GeV}$ so that its charge must be 
at least $10^{13}$ for it to evaporate for $10^{-12}$ s.
However, from Fig. \ref{pot3evsq} we see that scalar decays 
begin at about $Q\sim 10^{24}$. The lifetime of a typical Q-ball with 
charge of the order of $\sim (M/\textrm{(GeV)})^2$ is then 
sufficiently long 
for it to survive past the EW phase transition. On the other hand
a Q-ball with charge $\sim 10^{24}$ will evaporate in
$\sim 10^{23}\ \textrm{(GeV)}^{-1} \sim 10^{-2}$ s so that
Q-balls are no longer present in the universe.



We can now estimate the emitted power in Q-ball evaporation. 
Writing
\be{qwatts}
W=\dder{E}{t}=\dd{E}{Q}\dder{Q}{t}
\ee
so that we see that $\dd{E}{Q}$ must be considered as $E$ decreases. 
From
the Figures \ref{pot1evsq}, \ref{pot2evsq} and \ref{pot3evsq}
we can see that in the stable regime, $\dd{E}{Q}<m$ in all the 
three cases. The emitted power then increases until scalar decays 
begin. 
Especially in the case of potential 
$U_2(\phi)$ the emitted power can be quite large. However,
it must be kept in mind that this calculation is based on a 
semi classical analysis and very small Q-balls should
be analyzed quantum mechanically.

In analyzing the decay of Q-balls in the early universe 
one should also consider dissociation
and dissolution \cite{enqvistdm}. These are due to the presence
of a thermal background in the early universe. To evade the dissociation
and dissolution of typical Q-balls produced from an AD-condensate, an upper
bound for the reheating temperature of $\sim 10^{3-5}$ GeV 
has been derived \cite{enqvistdm}.

\section{Conclusions}
Q-ball properties have been studied using both analytical and numerical
methods.
Analytically we have examined the stability and charge of Q-balls 
in the limit where $\w\rightarrow m$.
In this thick wall limit we have derived conditions for stable Q-balls
to exist in a potential that can be expressed as powers of $\phi$ in the
limit of small $\phi$. It was shown that stability depends
on the dimension of the theory as well as on the power of the
next to leading term in the small $\phi$ limit. 
These results confirm previous results and present a method of
analyzing the stability of Q-balls when $\w\rightarrow m$.

Three different types of potentials have been studied numerically.
These calculations show how stability in the
thick-wall limit is strongly dependent on the exact form 
of the scalar potential. Numerical examinations show different types
of behaviour in the charge versus energy plot for all of the 
three potentials.
Furthermore, numerical work agrees with the analytical 
results.
In studying the potential resulting from a MSSM D-flat direction
with supergravity induced SUSY breaking we note that at least 
for the studied parameter values, the stability of Q-balls 
requires that $Q\sim 10^{24}$. This quite a large number
shows that one should be careful
to check whether the Q-balls formed in the early universe
will survive scalar decays for a given set of parameters
if Q-balls are to be cosmologically significant.

The evaporation of Q-balls was considered for a thin-walled Q-ball
and for realistic Q-ball profiles. We have 
recreated a previous result calculated for a semi-infinite ball.
For realistic Q-ball profiles it was shown that
as Q-balls lose their charge due to evaporation they evaporate
at a faster rate. Evaporation then progresses at an accelerating rate
until scalar decays begin. The emitted power also increases 
with decreasing charge so that Q-balls disappear in a burst of 
particles. Cosmologically
this means that if the universe was dominated by Q-balls at some point 
in its evolution,
there will be some inhomogeneity introduced by the evaporation process.
For example, if the baryon asymmetry or dark matter was
stored in Q-balls, inhomogeneity will be introduced
as the Q-balls evaporate.
This inhomogeneity could be cosmologically significant and possibly 
observable.

\newpage

\end{document}